\documentclass[twocolumn,showpacs,pra,amssymb,amsmath,floatfix,eqsecnum]{revtex4}

\usepackage{dcolumn}
\usepackage{amsmath,amssymb}
\usepackage{bm}
\usepackage{epsfig}

\begin{document}

  \title{Critical behavior of weakly interacting  bosons: A functional renormalization group
  approach}

  \author{Nils Hasselmann, Sascha Ledowski, and Peter Kopietz}

  \affiliation{Institut f\"{u}r Theoretische Physik, Universit\"{a}t
    Frankfurt, Robert-Mayer-Strasse 8, 60054 Frankfurt, Germany
}
  \date{\today}

  \begin{abstract}
  We present a detailed investigation of
the momentum-dependent self-energy $\Sigma ( k )$ at zero frequency
of weakly interacting bosons at the critical temperature $T_c$ of
Bose-Einstein condensation in dimensions $ 3 \leq D < 4$. Applying
the functional renormalization group, we calculate the
 universal scaling
function for the self-energy at zero frequency but at all wave vectors
within an approximation which truncates the flow equations of the
irreducible vertices at the four-point level. The self-energy
interpolates between the critical regime $ k \ll k_c$ and the
short-wavelength regime $ k \gg k_c$, where $k_c$ is the crossover
scale. In the critical regime, the self-energy correctly approaches
the asymptotic behavior $\Sigma ( k ) \propto k^{2 - \eta}$, and in
the short-wavelength regime  the behavior is $\Sigma ( k ) \propto
k^{2 (D-3)} $ in $D> 3$. In $D=3$, we recover the logarithmic
divergence $\Sigma ( k ) \propto \ln (k/k_c) $ encountered in
perturbation theory. Our approach yields the crossover scale  $k_c$
as well as a reasonable estimate for the critical exponent $\eta$ in
$D=3$. From our scaling function we find for the interaction-induced
shift in $T_c$ in three dimensions, $ \Delta T_c / T_c = 1.23 a
n^{1/3}$, where $a$ is the s-wave scattering length and $n$ is the
density, in excellent agreement with other approaches. We also
discuss the flow of marginal parameters in $D=3$ and extend our
truncation scheme of the renormalization group equations by
including the six- and
 eight-point vertex, which yields an improved
estimate for the anomalous dimension $\eta\approx 0.0513$. We further
calculate the constant $\lim_{k\to 0}\Sigma(k)/k^{2-\eta}$ and
find good agreement with recent Monte-Carlo data.

  \end{abstract}
\pacs{03.75.Hh, 05.30.Jp, 05.70.Jk}

  \maketitle

\section{Introduction}
\label{introduction} The physics of  weakly interacting bosons has
seen a revival thanks to the improved experimental technique of atom
traps which allow for a detailed study of Bose-Einstein condensation
in a controlled environment. Besides the theoretical effort invested
into studies of harmonically trapped bosons (for a review, see
Ref.~\cite{Dalfovo99}), this also generated  renewed interest
into the behavior of homogeneous Bose gases and led to some new
insights 
\cite{Andersen04}). Many aspects of
weakly interacting Bose gases have been understood for quite some
time now, such as the universality class $O(2)$ of the condensation
transition along with an accurate knowledge of critical exponents.
However, the universality of weakly interacting bosons at the
critical point is not limited to critical exponents since a weakly
interacting Bose gas behaves universal at all length scales 
larger than the thermal de-Broglie wavelength $\lambda_{\rm th}$
\cite{Baym99,Prokofev04}.
The origin of this extended universality is that at large length scales
quantum fluctuations become unimportant so that this
regime of the Bose gas is completely described by a classical
$\phi^4$-model \cite{Baym01}. In this work we present a detailed
study of the momentum-dependence of the self-energy at the
critical point of Bose-Einstein condensation at zero frequency,
using the functional renormalization group formalism in the form
introduced by Wetterich \cite{Wetterich93} and by Morris \cite
{Morris94}.
 Some results of this manuscript were already presented in a brief
form \cite{Ledowski04}. Here we give a detailed account of the
calculation and further include an extensive treatment of marginal
terms.

In the limit of weak interactions, parametrized by the s-wave
scattering length $a$, the self-energy is universal not only in the
limit of small wave vectors, ${\bf k}\to 0$, where it is
proportional to $k^{2-\eta}$ with
a finite anomalous dimension $\eta$. It remains
universal, in the sense that it can be written in a scaling form
independent of $a$,
 up to momenta which only need to be
small compared to 
$\lambda_{\rm th}^{-1}$. At some crossover scale $k_c$ the zero frequency
self-energy leaves the anomalous scaling regime and enters the
perturbative regime, where the self-energy correction to the dispersion becomes
negligible compared with the bare dispersion 
\begin{equation}
\epsilon_{\bf
k}=\rho_0 {\bf k}^2 \; , \hspace{.5cm} 
\mbox{with \ \ $\rho_0 =\hbar^2/2m$} \; .
\end{equation}
Here $m$ is the
bare mass. At finite temperatures and $D<4$, a 
perturbative calculation of the
self-energy is ultraviolet
(UV) divergent. This divergence can be addressed by using
the thermal de-Broglie length as an UV cutoff.
In $D=3$, however, the perturbative regime 
$k_c<k<2 \pi
/\lambda_{\rm th}$ remains non-trivial even in
presence of a UV cutoff, since additional logarithmic
infra-red (IR) divergences appear.
As discussed in detail by Baym
{\it et al.} \cite{Baym01}, the IR divergence cannot be treated in an
ad-hoc manner by introducing an IR cutoff, 
since doing so introduces an artificial
additional scale which directly enters quantities which should be
universal, such as the interaction induced shift of the critical
temperature. The IR divergence can be removed by a re-summation
using a variety of standard many-body techniques, e.~g.~bubble- or
ladder-summation and/or self-consistent approaches \cite{Baym01}.
However, these methods are uncontrolled in the critical regime
where one is faced with a strong-coupling
problem. Renormalization group (RG) techniques are expected to
perform better and several authors have applied RG techniques to
investigate the IR behavior of weakly interacting bosons
\cite{Bijlsma96,Andersen99,Pistolesi04}, though no attempt was made
to calculate the momentum dependence of the self-energy. Note that
standard field theoretical RG is confined to the critical regime
$k\ll k_c$. In fact, even the scale $k_c$ cannot be obtained within such
an approach. 
To interpolate between the critical and the short wavelength
regime, 
functional RG techniques \cite{Wetterich93,Morris94} are a natural choice, 
since they
track the flow of complete vertex functions rather than just a
small number of coupling parameters. Note that the functional RG approach
includes a priori also terms which are irrelevant according to their
scaling behavior, but which are important for a correct description of the
perturbative $k_c\ll k$ regime. In this work we present in detail a
functional RG approach to this problem and calculate the universal self-energy
to leading order in the small parameter
$an^{1/3}$ for all wave vectors ($n$ is the boson density).
Since the critical
behavior at $T_c$ is classical, we will focus 
only on the self-energy $\Sigma({\bf k}, i\omega_n=0)$ at vanishing
Matsubara frequency.

We begin with the general form of the functional RG flow equations for the
irreducible vertices up to the four-point vertex in
Sec.~\ref{SECflow}, where we still retain all frequencies. Our
approach is based on the sharp-cutoff version of the functional RG
\cite{Wetterich93,Morris94,Kopietz01,Busche01}. In
Sec.~\ref{SECcla}, we turn to the effective classical field theory
and rewrite the problem in a notation appropriate to the classical
limit. In Sec.~\ref{SECcla1}  the relevant and marginal 
parameters are classified in
dimensions $3< D<4$ and their functional flow equations are stated
in Sec.~\ref{SECcla2}.

Sec.~\ref{SECself} contains the central part of our work, the
calculation of the self-energy. We first re-examine perturbation
theory and its divergence in $D=3$ before we turn to the calculation
of the self-energy within the functional RG formalism. This formalism allows
us to calculate the scaling function $\sigma(x)$ characterizing the
zero frequency self-energy at the critical point. We define
$\sigma(x)$ by
\begin{eqnarray}
  \sigma(x)&=&(\rho_0 k_c^2)^{-1} [\Sigma(k_c x)-\Sigma(0)] \; ,
  \label{eq:sigmadef}
\end{eqnarray}
such that the crossover occurs at $x=1$. Here 
$\Sigma(k)=\Sigma(k, i\omega_n=0)$ is the
exact zero frequency self-energy at the critical temperature.
Standard field theoretical RG can only describe the asymptotic limit
$x\to 0$, where $\sigma(x)\propto x^{2-\eta}$.

Our approach is based on a truncation of the exact
hierarchy of functional RG flow
equations at the four-point vertex, i.~e.~we ignore six-point and
higher order  vertices. 
 While in this approximation the flow of
marginal terms is not consistently described, the resulting
approximation for the four-point vertex does in fact include
marginal and infinitely many irrelevant terms. 
The resulting flow equation for the self-energy can then
be solved and we derive the complete momentum dependence for the
self-energy in $3\le D<4$, see Eqs.~(\ref{eq:sigmares}, \ref{eq:FDdef})
below,
which constitute the central result of this work. A numerical
evaluation of Eqs.~(\ref{eq:sigmares}, \ref{eq:FDdef}) in $D=3$ is
also presented. We use $\sigma(x)$  to calculate the shift of the
critical temperature $T_c$ of the condensation transition in
Sec.~\ref{SECshift}. In $D=3$ we obtain $\Delta T_c/T_c =1.23 \ a n^{1/3}$
to lowest order in
$a n^{1/3}$,
in good
agreement with recent numerical investigations \cite{MC01a,MC01b}
and other analytical results \cite{Kleinert03,Kastening03} (for a
recent review on this topic see \cite{Andersen04}).

In Sec.~\ref{sec:self3}, we improve upon the truncation of the flow
equation and account for the coupling parameters which become marginal in
$D=3$. Three additional parameters must be taken into account, two of which are
associated with the linear momentum dependence of the four-point
vertex and one which describes the momentum-independent part of the
six-point vertex. We discuss different truncation schemes of the
flow equations and show that the inclusion of marginal terms provide
an improvement for the fixed point value of the
anomalous dimension.

In Sec.~\ref{sec:sigmamarg} we calculate $\sigma(x)$ in $D=3$
including the
marginal terms of the four-point vertex but ignoring irrelevant
terms.
We demonstrate how this truncation fails in the large $x$
regime, where it predicts incorrectly $\sigma(x)\propto x$ which in
turn would predict $\Delta T_c/T_c \propto a n^{1/3} \ln (a
n^{1/3})$. On the other hand, in the critical regime irrelevant
terms only lead to a renormalization of marginal and relevant ones
and $\sigma(x)$ is well described by a theory were irrelevant terms
are not included. We use this approach to express the prefactor
$A_3$ of the anomalous scaling term, $\sigma(x)\approx A_3
x^{2-\eta}$, as a function of $\eta$ for $D=3$. Our result
is in good agreement with recent Monte Carlo results
\cite{Prokofev04}. Finally, in Sec.~\ref{sec:conclusions} we
summarize and conclude this work.

\section{Functional RG flow equations for bosons}
\label{SECflow}

Our starting point is a standard
effective action describing free bosons with a two-particle
interaction, which is local at the bare level. The action with
an UV cutoff $\Lambda_0$ (to be specified later)
is of the form
 \begin{eqnarray}
 S_{\Lambda_0} \{ \bar{\psi}, \psi \} & = & S^0_{\Lambda_0} \{ \bar{\psi}, \psi \} +
 S^{\rm int}_{\Lambda_0} \{ \bar{\psi} , \psi \}
 \;  ,
 \end{eqnarray}
where the non-interacting  part is given by
 \begin{eqnarray}
 S^0_{\Lambda_0} \{ \bar{\psi} , \psi \}
 &=& \int_{ K } \Theta ( \Lambda_0 -
  | {\bf{k}} |  )
  \nonumber \\ & &
  \hspace{-10mm} \times \
 [ - i \omega_n + \epsilon_{ {\bf{k}} } -
 \mu + \Sigma (0, i0) ] \bar{\psi}_K \psi_{K}
 \; ,
 \label{eq:S0def}
 \end{eqnarray}
with $\Theta(x>0)=1$ and $\Theta(x<0)=0$.
The interaction part is given by
 \begin{eqnarray}
\hspace{-8mm}
  S^{\rm int}_{\Lambda_0} \{ \bar{\psi} , \psi \}
& = &
 \nonumber
 \\
 & & \hspace{-18mm}
 \int_{ K }
  \Theta ( \Lambda_0 -  | {\bf{k}} | )
 [ \Sigma_{\Lambda_0} (K)  -
 \Sigma (0 , i0) ] \bar{\psi}_{K}  \psi_{K}
 \nonumber
 \\
&  & \hspace{-18mm}
 + \frac{1}{ (2 !)^2} \int_{K_1^{\prime}}
 \int_{ K_2^{\prime} } \int_{K_2} \int_{K_1}
 \delta_{ K_1^{\prime} + K_2^{\prime} , K_2 + K_1}
 \nonumber
 \\
 & & \hspace{-18mm} \times \;
 \Gamma_{\Lambda_0}^{(4)} ( K_1^{\prime} , K_2^{\prime} ; K_2 , K_1 )
 \bar{\psi}_{ K_1^{\prime} }
  \bar{\psi}_{ K_2^{\prime} }
  \psi_{K_2} \psi_{K_1}
 + \ldots
  \; ,
 \label{eq:Sint}
 \end{eqnarray}
where the ellipsis denotes three-body and higher order interactions,
which we ignore at the bare level.
Here  $\psi_K$ is a complex  bosonic field.
We use the notation $K = ( {\bf{k}} , i \omega_n )$,
 $
 \int_K  =  (\beta V)^{-1} \sum_{ {\bf{k}} , \omega_n }
 $,
and
$ \delta_{ K  K^{\prime} }  =  \beta V
 \delta_{ {\bf{k}}  {\bf{k}}^{\prime} }
 \delta_{ \omega_n  \omega_{n^{\prime}} }$,
where $\beta$ is the inverse temperature, $V$ is the volume, and
$\omega_n = 2 \pi n T$ are bosonic Matsubara frequencies. In
Eq.~(\ref{eq:S0def}) we have included the exact self-energy at
vanishing momenta and frequencies as a counterterm in the definition
of the free action. Throughout this work, we shall work at
temperatures $T\ge T_c$ such that the $U(1)$ symmetry is not broken.
The generating functional of the one-particle irreducible $n$-point
vertices $\Gamma^{(2n)}_{\Lambda} ( K_1^{\prime} , \ldots ,
K_n^{\prime} , K_n , \ldots , K_1  )$ of the theory with IR cutoff
$\Lambda$ can be expanded in terms of the fields $\phi_K = \langle
\psi_K \rangle$ as follows
 \begin{eqnarray}
 \Gamma_{\Lambda} \{ \bar{\phi} , \phi \}  & = & \sum_{n=0}^{\infty}
 \frac{1}{ (n! )^2} \prod_{i,j = 1}^{n} \int_{K^{\prime}_i} \int_{K_j}
\delta_{ K_1^{\prime} + \ldots + K_n^{\prime} , K_n + \ldots + K_1 }
 \nonumber
 \\
 &  & \hspace{-7mm} \times
 \Gamma^{(2n)}_{\Lambda} ( \left\{ K_i^{\prime} , K_j \right\} )
 \bar{\phi}_{ K_1^{\prime} } \cdots \bar{\phi}_{ K_n^{\prime}}
 {\phi}_{ K_n} \cdots {\phi}_{ K_1}
 \; .
 \label{eq:generatingfuncdef}
 \end{eqnarray}
The functional RG flow equations of the first few irreducible vertices (up to
the six-point vertex) for non-relativistic fermionic and bosonic
many-body systems can be found in Ref.~\cite{Busche01}. We
summarize below the flow equations relevant for bosons up to the
four-point vertex. The flow equations for the six-point vertex and
some terms of the flow of the eight-point vertex are given in
Appendix \ref{AppMarg}.

 \subsection{Free energy}
For completeness, we list here
the flow equation for the free energy $\Gamma_\Lambda^{(0)}$, 
although below we shall not
discuss it further. The flow equation is given by
 \begin{eqnarray}
 \partial_{\Lambda} \Gamma^{(0)}_{\Lambda} &=&
   V  \int_K \delta ( \Lambda - | {\bf{k}} |  )  \nonumber \\
&& \hspace{-1.3cm} \times \ln \left[ \frac{  i \omega_n -
\epsilon_{\bf{k}}
 +   \mu - \Sigma (0 , i0) }{ i \omega_n - \epsilon_{\bf{k}} + \mu
 - \Sigma (0,i0) - \Gamma_{\Lambda}^{(2)} ( K ) }
 \right]
 \; ,
 \label{eq:flowGamma0}
 \end{eqnarray}
where $\Gamma^{(2)}_{\Lambda}(K)$ is the two-point vertex, defined
by
\begin{equation}
 \Gamma^{(2)}_{\Lambda} ( K ) =  \Sigma_{\Lambda} ( K ) - \Sigma (0 , i
 0) \; .
 \end{equation}
\subsection{Two-point vertex}
The main interest of the work is the flow equation of the two point
vertex, which we will relate in Sec.~\ref{SECself} to the zero
energy scaling function of the self-energy, Eq.~(\ref{eq:sigmadef}).
The flow equation for the two-point vertex is given by
 \begin{eqnarray}
\hspace{-8mm}
 \partial_\Lambda \Gamma^{(2)}_{\Lambda} ( K ) & = &
  \int_{ K^{\prime}}
 \dot{G}_{\Lambda} ( K^{\prime} )
 \Gamma^{(4)}_{\Lambda} ( K , K^{\prime} ; K^{\prime} , K )
 \; ,
 \label{eq:flowGamma2}
 \end{eqnarray}
where
 \begin{equation}
 \dot{G}_{\Lambda} ( K ) =
 \frac{ \delta ( \Lambda -  | {\bf{k}} |  )}{
 i \omega_{n} - \epsilon_{ {\bf{k}} } + \mu
 - \Sigma (0, i0) - \Gamma^{(2)}_{\Lambda} ( K ) }
 \label{eq:dotGk}
   \;
 \end{equation}
is the cutoff dependent  single scale propagator, with support only at 
$|{\bf k}|=\Lambda$.
As is evident from Eq.~(\ref{eq:flowGamma2}), the flow of the two
point vertex depends on the properties of the four-point vertex at
finite wave vectors and we thus
need to derive the momentum-dependent flow of the four-point
vertex.

\subsection{Four-point vertex}
\begin{figure}
\epsfxsize8.5cm
\hspace{5mm} \epsfbox{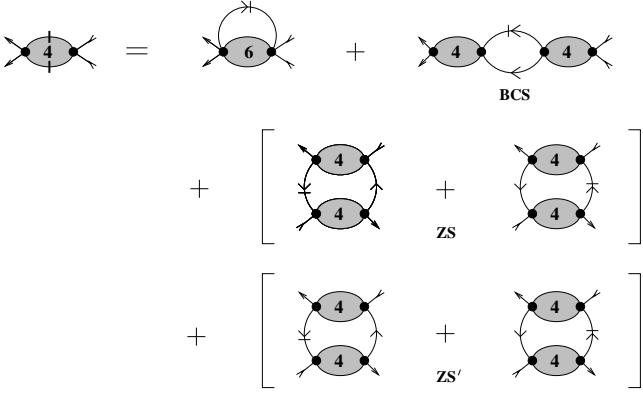} \vspace{5mm} \caption{
Diagrammatic representation of the flow equation for the four-point
vertex, see Eq.~(\ref{eq:flowGamma4}). We adapt the notation BCS, ZS
and ZS$^\prime$ for the diagrams from the usual fermionic language
\cite{Shankar94},
even though this notation does not imply a physical correspondence.}
\label{fig:FourPoint}
\end{figure}

The flow equation for the irreducible four-point vertex is responsible for
the cross-over from the weak coupling regime at small scales to the
critical regime at large scales. The flow is given by
\begin{widetext}
 \begin{eqnarray}
 \partial_\Lambda \Gamma^{(4)}_{\Lambda} ( K_1^{\prime} , K_2^{\prime} ; K_2 , K_1 ) & =&
    \int_{ K}  \dot{G}_{\Lambda} ( K )
 \Gamma^{(6)}_{\Lambda} ( K_1^{\prime} , K_2^{\prime} , K  ; K , K_2 , K_1 )
 \nonumber
 \\
 & & \hspace{-38mm} + \int_K
  \left[  \dot{G}_{\Lambda} ( K )
 G_{\Lambda } ( K^{\prime} )
 \Gamma^{(4)}_{\Lambda} ( K_1^{\prime} , K_2^{\prime} ; K^{\prime} , K )
 \Gamma^{(4)}_{\Lambda} ( K , K^{\prime} , K_2 , K_1 )
 \right]_{K^{\prime} = K_1 + K_2 - K }
 \nonumber
 \\
 & & \hspace{-38mm} +
 \int_K \left[
\bigl[
\dot{G}_{\Lambda} ( K )
 G_{\Lambda } ( K^{\prime} )
 +
 G_{\Lambda} ( K )   \dot{G}_{\Lambda} ( K^{\prime} )
 \bigr]
 \Gamma^{(4)}_{\Lambda} ( K_1^{\prime} ,  K^{\prime} ;  K , K_1 )
 \Gamma^{(4)}_{\Lambda} ( K_2^{\prime} , K ; K^{\prime} , K_2 )
 \right]_{K^{\prime} =  K_1 - K_1^{\prime} + K }
 \nonumber
 \\
 & & \hspace{-38mm}   +
  \int_K \left[
\bigl[
\dot{G}_{\Lambda} ( K )
 G_{\Lambda } ( K^{\prime} )
 +
 G_{\Lambda} ( K )   \dot{G}_{\Lambda} ( K^{\prime} )
 \bigr]
 \Gamma^{(4)}_{\Lambda} ( K_2^{\prime} ,  K^{\prime} ;  K , K_1 )
 \Gamma^{(4)}_{\Lambda} ( K_1^{\prime} , K ; K^{\prime} , K_2 )
 \right]_{K^{\prime} =  K_1 - K_2^{\prime} + K}
 \; ,
 \label{eq:flowGamma4}
 \end{eqnarray}
\end{widetext}
where  $\Gamma^{(6)}_\Lambda$ is the six-point vertex and
 \begin{equation}
 G_{\Lambda } ( K ) = \frac{ \Theta ( \Lambda < | {\bf{k}} |
 < \Lambda_0 )}{
 i \omega_{n} - \epsilon_{ {\bf{k}} } + \mu
 - \Sigma (0 , i0) - \Gamma^{(2)}_{\Lambda} ( K ) }
 \;
 \end{equation}
is the cutoff regularised propagator. The notation 
$\Theta ( \Lambda < | {\bf{k}} | < \Lambda_0 )$ is shorthand for
$\Theta(| {\bf{k}} |-\Lambda)-\Theta(| {\bf{k}} |-\Lambda_0)$.
This equation is shown
graphically in Fig.~\ref{fig:FourPoint}. For the most part of this
work, we shall in fact ignore the contribution from the
six-point vertex to
this flow and work solely with the two- and four-point vertex. To
calculate fixed-point properties, we will however also include the
six-point vertex in Sec.~\ref{sec:self3}.

\section{Effective classical field theory}
\label{SECcla} To discuss the classical critical behavior, it is
sufficient to retain only the zero Matsubara frequency part of all
vertices. In principle, all fields with non-zero Matsubara
frequencies can be integrated out using the flow equations given
above, which leads to finite renormalizations of the parameters
appearing in the effective classical theory \cite{Arnold01}. The
effective UV cutoff of the classical theory is determined by the
thermal de Broglie wavelength,
 \begin{equation}
 \Lambda_0 = \frac{2 \pi}{\lambda_{\rm th}} \; \; , \; \;
 \lambda_{\rm th} = \frac{ h}{\sqrt{ 2 \pi m  T}} =
 \sqrt{ \frac{ 2 \pi \hbar^2}{ m T  } }
 \;.
 \end{equation}
In $D=3$ the initial value of the four-point vertex can be
parameterized in terms of the two-body scattering length
$a$,
 \begin{equation}
 \Gamma^{(4)}_{\Lambda_0} ( 0,0;0,0) = \frac{ 8 \pi \hbar^2 a}{m} = 16 \pi \rho_0 a
 \; .
 \label{eq:scatteringlength}
 \end{equation}

\subsection{Functional RG flow equations for the rescaled vertices}
\label{SECcla1} We assume that the finite renormalizations due to
non-zero Matsubara frequencies are implicitly  taken into account
via the initial conditions at scale $\Lambda = \Lambda_0$ of the
effective classical theory. As discussed in detail in
Ref.~\cite{Baym01}, to calculate the linear shift of the
critical temperature (see Sec.~\ref{SECshift}), one can in fact
ignore the renormalization of the classical sector by finite
Matsubara frequencies altogether, since this only leads to
corrections in $T_c$ of order $a^2$. It is convenient to write the
single-particle Green's function at zero Matsubara frequency in the
following scaling form
 \begin{equation}
\label{greenscal}
 G_{\Lambda } ( {\bf{k}} , i0 ) =
 - \frac{Z_l}{ \rho_0 \Lambda^2} \tilde{G}_l (  {\bf{k}}  \Lambda )
 \; ,
 \end{equation}
 \begin{equation}
 \tilde{G}_l (  {\bf{q}}   ) = \frac{ \Theta ( 1 < | {\bf{q}} | < e^l )}{
 R_l (  {\bf{q}}  )}
 \; ,
 \end{equation}
were the minus sign is introduced in Eq.~(\ref{greenscal}) to arrive
at the usual definition of the classical Green's function.
Here $l = - \ln ( \Lambda / \Lambda_0 )$ is the
logarithmic flow parameter, and
the inverse dimensionless propagator is
 \begin{equation}
 R_l ( {\bf{q}} ) = Z_l {\bf{q}}^2 + \tilde{\Gamma}^{(2)}_l ( {\bf{q}} )
 + \frac{Z_l}{ \rho_0 \Lambda^2} [ \Sigma (0,i0) - \mu ] \; ,
 \end{equation}
with the dimensionless irreducible two-point vertex
 \begin{eqnarray}
   \label{2pointdimless}
 \tilde{\Gamma}^{(2)}_l ( {\bf{q}} ) & = &
  \frac{ Z_l}{\rho_0 \Lambda^2}
 \Gamma_{\Lambda}^{(2)} ( \Lambda {\bf{q}} , i 0 )
 \nonumber
 \\
 & = &
 \frac{ Z_l}{\rho_0 \Lambda^2} [ \Sigma_\Lambda ( \Lambda {\bf{q}} , i 0 )
 - \Sigma ( 0 , i 0) ]
 \; .
 \end{eqnarray}
 The classical wave-function normalization factor $Z_l$ is
given by
 \begin{equation}
 Z_l = 1 - \left.
\frac{ \partial \tilde{\Gamma}^{(2)}_l ( {\bf{q}} ) }{\partial q^2}
 \right|_{ q^2 = 0 }
 \; .
 \end{equation}

Similar to Eq.~(\ref{2pointdimless}), we define
the classical dimensionless higher order vertices for $n \geq 2$
by
 \begin{eqnarray}
   \label{rescalegeneral}
   &&
   \tilde{\Gamma}^{(2n)}_l ( {\bf{q}}_1^{\prime} , \ldots ,
   {\bf{q}}_n^{\prime} ; {\bf{q}}_n , \ldots , {\bf{q}}_1 ) \nonumber
   = ( K_D T )^{n-1}    \\ &&  \ \ \
   \times  \Lambda^{ D (n-1) - 2n }
   \left( {Z_l}/{\rho_0} \right)^n \Gamma_{\Lambda}^{(2n)} (
   \{   {\bf{k}}_i = \Lambda {\bf{q}}_i  ,  \omega_{n_i} =0  \}    )
   \; . 
\nonumber \\
 \end{eqnarray}
For later convenience we have included the numerical factor
$K_D = \Omega_D / ( 2 \pi )^D$ in the definition
of the vertices, where
$\Omega_D = 2 \pi^{D/2}/\Gamma ( D/2)$
 is the
surface area of the $D$-dimensional unit sphere.
The rescaled vertices satisfy functional flow equations of the form
 \begin{eqnarray}
   \label{eq:rescaledflowgeneral}
   &&
   \partial_l  \tilde{\Gamma}^{(2n)}_l (  \{ {\bf{q}}_i \} )
   = \nonumber
   \\ \nonumber &&
   \left[ 2n - D ( n-1 ) - n \eta_l -
     \sum_{i=1}^{2 n}
     {\bf{q}}_i \cdot \nabla_{ {\bf{q}}_i}
   \right]  \tilde{\Gamma}^{(2n)}_l (  \{ {\bf{q}}_i \} )
   \\  &&
   + \dot{{\Gamma}}^{(2n)}_l (  \{ {\bf{q}}_i \} )
   \; ,
 \end{eqnarray}
where
 \begin{equation}
 \eta_l = - \partial_l \ln Z_l
 \end{equation}
is the flowing anomalous dimension.
In particular, the rescaled two-point vertex satisfies
 \begin{equation}
 \partial_l \tilde{\Gamma}_l^{(2)} ( {\bf{q}} )
 = [ 2 - \eta_l - {\bf{q}} \cdot \nabla_{ {\bf{q}} } ]
\tilde{\Gamma}_l^{(2)} ( {\bf{q}} ) + \dot{\Gamma}_l^{(2)} (
{\bf{q}} )
 \; ,
 \label{eq:flowtildeGamma2}
 \end{equation}
where
 \begin{eqnarray}
 \dot{\Gamma}_l^{(2)} ( {\bf{q}} )
 & = &
 \int_{\bf{q}^{\prime}}
 \dot{G}_l ( {\bf{q}}^{\prime} )
\tilde{\Gamma}^{(4)}_l ( {\bf{q}} , {\bf{q}}^{\prime} ;
 {\bf{q}}^{\prime} , {\bf{q}} )
 \; ,
 \label{eq:gammadot2def}
 \end{eqnarray}
with
 \begin{equation}
 \int_{\bf{q}^{\prime}}   =
\int \frac{ d^D q^{\prime}}{ \Omega_D}
 \; ,
\end{equation}
 and
 \begin{equation}
 \dot{G}_l ( {\bf{q}} ) =
 \frac{ \delta ( | {\bf{q}} | -1 )}{R_l ( {\bf{q}} )}
 \; .
 \end{equation}
The functional flow equation for the dimensionless four-point vertex follows
 from Eq.~(\ref{eq:rescaledflowgeneral}). It involves
the inhomogeneity
\begin{widetext}
 \begin{eqnarray}
 \dot{\Gamma}_{l}^{(4) }
( {\bf{q}}_1^{\prime} , {\bf{q}}_2^{\prime} ; {\bf{q}}_2 , {\bf{q}}_1 )
  & = &
\int_{\bf{q}} \dot{G}_l ( {\bf{q}} )
 \tilde{\Gamma}^{(6)}_l ( {\bf{q}}_1^\prime ,
{\bf{q}}_2^\prime , {\bf{q}} ; {\bf{q}} ,
 {\bf{q}}_2 , {\bf{q}}_1 )
 \nonumber
-  \int_{ {\bf{q}} }
 \left[ \dot{G}_l ( {\bf{q}} ) \tilde{G}_l  ( {\bf{q}}^{\prime} )
 \tilde{\Gamma}^{(4)}_{l} ( {\bf{q}}_1^{\prime} , {\bf{q}}_2^{\prime} ;
 {\bf{q}}^{\prime} , {\bf{q}} )
 \tilde{\Gamma}^{(4)}_{l} ( {\bf{q}} , {\bf{q}}^{\prime} ; {\bf{q}}_2 , {\bf{q}}_1 )
 \right]_{{\bf{q}}^{\prime} = {\bf{q}}_1 + {\bf{q}}_2 - {\bf{q}} }
 \nonumber
 \\
 & &
- \int_{\bf{q}}
 \Bigl[ \bigl[ \dot{G}_l ( {\bf{q}} ) \tilde{G}_l  ( {\bf{q}}^{\prime} )
 +\tilde{G}_l ( {\bf{q}} ) \dot{G}_l  ( {\bf{q}}^{\prime} )
 \bigr]
 \tilde{\Gamma}^{(4)}_{l} ( {\bf{q}}_1^{\prime} , {\bf{q}}^{\prime} ;
 {\bf{q}} , {\bf{q}}_1 )
 \tilde{\Gamma}^{(4)}_{l} ( {\bf{q}}_2^{\prime} , {\bf{q}} ; {\bf{q}}^{\prime} ,
{\bf{q}}_2 )
 \Bigr]_{{\bf{q}}^{\prime} =  {\bf{q}}_1 - {\bf{q}}_1^{\prime}  + {\bf{q}} }
 \nonumber
 \\
 & &
 - \int_{\bf{q}}
 \Bigl[ \bigl[
 \dot{G}_l ( {\bf{q}} ) \tilde{G}_l  ( {\bf{q}}^{\prime} )
 +\tilde{G}_l ( {\bf{q}} ) \dot{G}_l  ( {\bf{q}}^{\prime} )
\bigr]
 \tilde{\Gamma}^{(4)}_{l} ( {\bf{q}}_2^{\prime} , {\bf{q}}^{\prime} ;
 {\bf{q}} , {\bf{q}}_1 )
 \tilde{\Gamma}^{(4)}_{l} ( {\bf{q}}_1^{\prime} , {\bf{q}} ; {\bf{q}}^{\prime} ,
{\bf{q}}_2 )
 \Bigl]_{{\bf{q}}^{\prime} =  {\bf{q}}_1 - {\bf{q}}_2^{\prime}  +  {\bf{q}}  }
 \label{eq:fourpointscale}
 \; .
 \end{eqnarray}
\end{widetext}
The flow equations for the six-point and eight-point vertices also
have the form Eq.~(\ref{eq:rescaledflowgeneral}) where
$\dot{\Gamma}_{l}^{(6)}$ and $\dot{\Gamma}_{l}^{(8)}$ involve
various combinations of the two-, six-, four-, eight- and ten-point
vertices. All terms entering the inhomogeneity of the six-point
vertex and 
the terms needed for a calculation up to second order in
the relevant and marginal parameters of the inhomogeneity of the
eight-point 
can be found in Appendix
\ref{AppMarg}, see Eqs.~(\ref{eq:sixpoint},\ref{eq:eightpoint}).

\subsection{Classification of coupling parameters}
\label{SECcla2}
Although we are ultimately interested in the flow of vertex
functions, it is useful to first consider the flow of
marginal and relevant coupling parameters, since irrelevant
parameters become local functions of the
relevant and marginal ones at the fixed point \cite{Polchinski84}.
To properly organize the flow of
irrelevant terms it is thus necessary to know the flow
of the relevant and marginal parameters. We first
investigate here the relevant and marginal terms
for $3<D<4$, ignoring additional marginal terms in $D=3$.
\subsubsection{Relevant coupling parameters}
\label{sec:relevant}
In  $ 3 \leq  D < 4$ there are two relevant coupling parameters,
 \begin{equation}
r_l = \tilde{\Gamma}_l^{(2)} ( 0 )
 = \lim_{ {\bf{q}} \rightarrow 0}
\frac{  Z_l}{\rho_0 \Lambda^2} [ \Sigma_\Lambda ( \Lambda {\bf{q}} , i 0 )
 - \Sigma ( 0 , i 0) ]
 \; \; ,
 \end{equation}
with scaling dimension $+2$, and
 \begin{equation}
u_l = \tilde{\Gamma}_l^{(4)} ( 0 , 0 ; 0 , 0)
 \; \; ,
 \end{equation}
 with scaling dimension $ \epsilon = 4 - D$.
The exact flow equations of these parameters are
 \begin{equation}
 \partial_l r_l = ( 2 - \eta_l ) r_l +
 \dot{\Gamma}^{(2)}_l ( 0 )
 \; ,
 \label{eq:exactflowrl}
 \end{equation}
 \begin{equation}
 \partial_l u_l =
 ( 4 - D - 2 \eta_l ) u_l +  \dot{\Gamma}^{(4)}_l ( 0 , 0 ; 0, 0)
  \; .
 \label{eq:exactflowul}
 \end{equation}

\subsubsection{Marginal coupling parameter}
\label{subsec:marginal}

For $3 < D < 4$ the only marginal parameter is the
wave-function renormalization $Z_l$.
The exact flow equation is
 \begin{equation}
 \partial_l Z_l = - \eta_l Z_l
 \; ,
 \label{eq:etagammadot}
 \end{equation}
where the flowing anomalous dimension is
 \begin{eqnarray}
 \eta_l & = & \left. \frac{ \partial \dot{\Gamma}^{(2)}_l ( {\bf q} ) }{
 \partial q^2 } \right|_{ q^2 =0}
 \nonumber
 \\
 & = & \frac{1}{R_l ( 1 )}
  \frac{ \partial}{\partial q^2 }
\big<
        \tilde{\Gamma}_l^{(4)} ( {\bf{q}} , {\bf{\hat{q}^\prime}} ;
 {\bf{\hat{q}^\prime}} , {\bf{q}}  )
\big>_{\bf \hat{q}^\prime}
 \Big|_{ q^2 =0}
 \; ,
 \label{eq:anomalexplicit}
 \end{eqnarray}
and
\begin{eqnarray}
\big< \dots \big>_ {\bf \hat{q}}&=&\int\frac{d^{D} {\bf q}} {\Omega_D}
\delta(|{\bf q}|-1)\
 \dots
\end{eqnarray}
denotes the integral over the unit sphere. The surface area
$\Omega_D$ of the unit sphere was defined below
Eq.~(\ref{rescalegeneral}). In $D=3$, additional marginal coupling
parameters appear which we ignore in this section. We will discuss
them 
in Sec.~\ref{sec:self3}.

\subsubsection{The Wilson-Fisher fixed point close to $D=4$}

\begin{figure}
\epsfysize5.5cm
\epsfbox{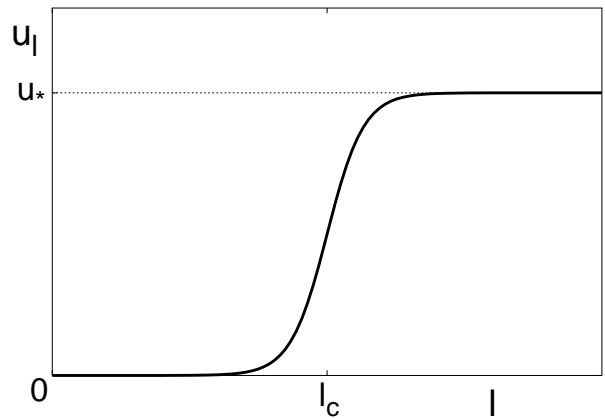} 
\caption{ Typical flow  of the
coupling $u_l$ as a function of $l$ for $u_0 \ll u_{\ast}$ and
$D=3$. } \label{fig:uflow}
\end{figure}

It is instructive to take a closer look  at the perturbative
one-loop RG flow equations for the relevant coupling parameters. If we retain
from the two-point vertex only the relevant and marginal part, we may
approximate
 \begin{equation}
 R_l ({\bf{q}} ) \approx r_l + {\bf{q}}^2
 \; ,
 \label{eq:Rlapprox}
 \end{equation}
so that
 \begin{equation}
 \tilde{G}_l ( {\bf{q}} ) \approx \frac{ \Theta ( 1 < | {\bf{q}} | < e^{l} )}{ r_l + {\bf{q}}^2 }
 \; ,
 \label{eq:Glapprox}
 \end{equation}
 \begin{equation}
 \dot{G}_l ( {\bf{q}} ) \approx \frac{ \delta ( | {\bf{q}} | -1)}
{ r_l + {\bf{q}}^2 }
 \; .
 \label{eq:dotGlapprox}
 \end{equation}
If in addition we retain only the relevant part of the
four-point vertex, we may replace
it by a momentum-independent constant,
 \begin{equation}
\tilde{\Gamma}_{l}^{ (4)}
 ( {\bf{q}}_1^{\prime} , {\bf{q}}_2^{\prime} ; {\bf{q}}_2 , {\bf{q}}_1 )
 \approx  \tilde{\Gamma}_{l}^{ (4)}
 ( 0 , 0 ; 0 , 0 ) \equiv u_l
 \; .
 \end{equation}
Within this approximation we obtain from
Eq.~(\ref{eq:gammadot2def}),
 \begin{equation}
 \dot{\Gamma}^{(2)}_l ( 0 ) \approx \frac{ u_l}{ 1 + r_l}
 \; ,
 \label{eq:dotgamma21}
 \end{equation}
and
Eq.~(\ref{eq:fourpointscale}) reduces to
 \begin{equation}
\dot{\Gamma}^{(4)}_l ( 0 , 0 ; 0, 0) \approx
 -  \frac{5 }{2}
\frac{ u_l^2}{ ( 1 + r_l )^2 }
  \; .
 \label{eq:dotgamma41}
 \end{equation}
Because the anomalous dimension is related to the
momentum dependence of the four-point vertex which is irrelevant
in $D > 3$, we may set $\eta_l \approx 0$ within this approximation.
Furthermore, on the critical trajectory
$r_l = O ( u_l )$ can also be ignored in Eq.~(\ref{eq:dotgamma41})
as long as $u_l$ remains
small compared with unity. Then we obtain, with  $\epsilon =  4-D$,
 \begin{equation}
 \partial_l u_l = \epsilon u_l - \frac{5}{2}   u_l^2 \; ,
 \label{eq:gflow}
 \end{equation}
 \begin{equation}
 \partial_l r_l = 2 r_l +
 { u_l}
 \label{eq:muflow}
 \; .
 \end{equation}
Eq.~(\ref{eq:gflow}) is easily solved. The solution can be written
in the form
 \begin{equation}
 \frac{u_l}{u_{\ast} }
 = 1 - \frac{1}{ e^{\epsilon ( l - l_c ) } +1 }
 = \frac{1}{ e^{ \epsilon ( l_c - l ) } +1 }
 \; ,
 \label{eq:Fermi}
 \end{equation}
where
\begin{equation}
 u_{\ast}
 = \frac{ 2}{5} \epsilon
 \end{equation}
is the value of $u_l$ at the RG fixed point, and the
logarithmic crossover scale is
 \begin{equation}
 l_c = \frac{1}{\epsilon} \ln \left( \frac{u_{\ast}}{u_0} -1
 \right) \approx
 \frac{1}{\epsilon} \ln \left( \frac{u_{\ast}}{u_0}
 \right)
 \; .
 \end{equation}
Throughout this work we assume
that $u_0 \ll u_{\ast}$, corresponding to weak bare interactions
between the bosons.
Note that the right-hand side of Eq.~(\ref{eq:Fermi}) is expressed in terms of
the Fermi function. If we think of $\epsilon$ as the inverse temperature
and $l_c$ as the  chemical potential, the
qualitative behavior of $u_l / u_{\ast}$ is clear:
within a narrow  (on the scale $l_c$)
interval of width $1/ \epsilon$ centered at $l = l_c$
the ratio $u_l / u_{\ast}$ raises from the small value
 $e^{ - \epsilon ( l_c - l ) }$ to a value close to unity,
see Fig. \ref{fig:uflow}.
Given the solution $u_l$ in Eq.~(\ref{eq:Fermi}),
the RG equation (\ref{eq:muflow}) for $r_l$
is easily solved,
 \begin{equation}
 r_l = - \frac{\epsilon}{5}
 + e^{2 l } \left[
 r_0 + \frac{\epsilon}{5} -  \frac{ 2 \epsilon}{5}
 \int_0^{l} d l^{\prime}
 \frac{ e^{- 2 l^{\prime}}}{ e^{ \epsilon ( l^{\prime} - l_c )} + 1}
 \right]
 \; .
 \label{finetune1}
 \end{equation}
To obtain a fixed point at $l\to \infty$, the intial value $r_0$
has to be fine tuned such that
 \begin{equation}
 r_0 =  - \frac{\epsilon}{5} + \frac{ 2 \epsilon}{5}
 \int_0^{\infty} d l^{\prime}
 \frac{ e^{- 2 l^{\prime}}}{ e^{ \epsilon ( l^{\prime} - l_c )} + 1 }
 \; .
 \label{eq:r0fix}
 \end{equation}
Then we may write
 \begin{equation}
 r_l = r_{\ast} +   u_{\ast}
 \int_l^{\infty} d l^{\prime} \frac{ e^{- 2 ( l^{\prime} - l )}}{
 e^{\epsilon ( l^{\prime} - l_c )} +1 }
 \; ,
 \end{equation}
where
 \begin{equation}
 r_{\ast} = - \frac{u_{\ast}}{2}
 = - \frac{\epsilon}{5}
 \; .
 \label{WilsonFisherFP}
 \end{equation}
 In Sec.~\ref{SECself} we show that,
up to a numerical factor of the order of unity,
the momentum scale $k_c = \Lambda_0 e^{- l_c}$
associated with the logarithmic scale factor $l_c$, i.~e.
 \begin{equation}
 k_c =  \Lambda_0 \left[
 \frac{u_{\ast}}{u_0} -1
 \right]^{ - 1/ \epsilon } \approx
\Lambda_0 \left( \frac{ u_0 }{ u_{\ast} } \right)^{1/ \epsilon}
 \; ,
 \label{eq:Lambdac}
 \end{equation}
can be  identified with  the  crossover scale
where the  critical $k^{ 2 - \eta}$-form
of the energy dispersion begins to emerge.

\section{The self-energy at the critical point of Bose-Einstein condensation}
\label{SECself}

\subsection{Second  order perturbation theory}
\label{sec:selfenergypert}

\begin{figure}
\epsfysize4.8cm
\epsfbox{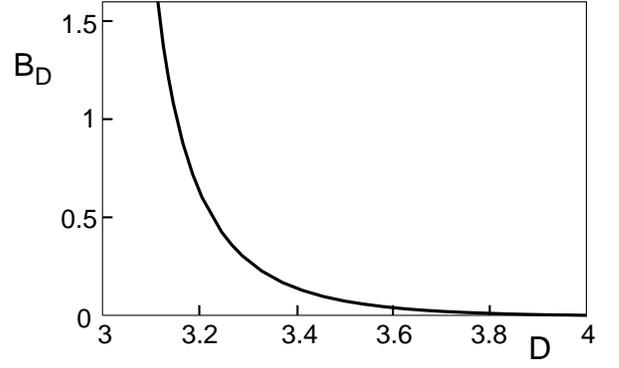}
\caption{
Graph of the coefficient $B_D$
defined in Eqs.~(\ref{eq:BDres}) and (\ref{eq:Iprime})
in dimensions $3 < D \leq 4$.
}
\label{fig:BD}
\end{figure}

To begin with, let us attempt to calculate the self-energy by means
of straightforward second order perturbation theory \cite{Baym01},
which yields in a continuum model with UV cutoff $\Lambda_0$
 \begin{eqnarray}
  \Sigma ( {\bf{k}} ) - \Sigma ( 0 )
 &=& - T \frac{3}{2} \Gamma_0^2 \int \frac{d^D p}{ ( 2 \pi )^D}
 \chi_0( {\bf{p}} ) \\ & &\times
 \left[ \frac{ \Theta ( \Lambda_0 - | {\bf{p}} + {\bf{k}} | ) }{ \epsilon_{ {\bf{p}} + {\bf{k}} } }
  - \frac{ \Theta ( \Lambda_0 - |  {\bf{p}} | ) }{ \epsilon_{ {\bf{p}} } }
 \right]
 \; , \nonumber
 \label{eq:pertsigma}
 \end{eqnarray}
 where $\Gamma_0 = \Gamma_{\Lambda_0}^{(4)} (0,0; 0,0)$ is the bare vertex, and
 \begin{equation}
  \chi_0 ( {\bf{p}} ) = T  \int \frac{d^D p^{\prime}}{ ( 2 \pi )^D}
 \frac{ \Theta ( \Lambda_0 - | {\bf{p}}^{\prime} | )
 \Theta ( \Lambda_0 - | {\bf{p}}^{\prime}  + {\bf{p}} | )}{
 \epsilon_{  {\bf{p}}^{\prime}} \epsilon_{ {\bf{p}}^{\prime} + {\bf{p}} } }
 \; .
 \end{equation}
With $ \epsilon_{\bf{p}} = \rho_0 {\bf{p}}^2$
we obtain for $ \Lambda_0 \rightarrow \infty$,
 \begin{equation}
 \chi_0 ( {\bf{p}} ) =   K_D K_D^{\prime}  \frac{T }{ \rho_0^2 p^{4 -D} }
 \label{eq:chi0res}
\end{equation}
where
 \begin{eqnarray}
 K_D^{\prime} & = & \int_0^1 dx [ x (1-x)]^{ (D-4)/2}
 \int_0^{\infty} dy \frac{ y^{D-1}}{ ( y^2 +1 )^2 }
 \nonumber
 \\
 & = & 2^{3-D} \frac{ \sqrt{\pi} \Gamma ( \frac{D}{2} -1 )}{\Gamma (
 \frac{D-1}{2} )}  \frac{ \frac{\pi}{4} (D-2)}{ \sin ( \frac{\pi}{2}
 (D-2) }
 \; .
 \label{eq:IDdef}
 \end{eqnarray}
Note that $K_D^{\prime}$ is finite for $2 < D < 4$; in particular,
$K_3^{\prime} = \pi^2 /4$. Substituting Eq.~(\ref{eq:chi0res}) into
Eq.~(\ref{eq:pertsigma}) and taking the limit $\Lambda_0 \rightarrow
\infty$, we can scale out the $k$-dependence and obtain for $D< 3
<4$, using the definitions $u_0 =  K_D T \Lambda_0^{- \epsilon}
\rho_0^{-2} \Gamma_0$ and $ k_c = \Lambda_0 ( u_0 / u_{\ast} )^{1/
\epsilon}$,
 \begin{equation}
  \rho_0^{-1} [ \Sigma ( {\bf{k}} ) - \Sigma ( 0 ) ] = B_D k_c^{2 \epsilon} k^{2 - 2 \epsilon}
 \; ,
 \label{eq:selfpert2}
 \end{equation}
where after an integration by parts the coefficient $B_D$ can be cast into the following form,
 \begin{equation}
 B_D =  \frac{3  u_{\ast}^2  }{4 (D-3)} K_D^{\prime} {I}_D
 \; ,
 \label{eq:BDres}
 \end{equation}
 \begin{eqnarray}
 {I}_D & = & \frac{\Omega_{D-1}}{\Omega_D} \int_0^{\pi}
 d \vartheta ( \sin \vartheta )^{D-2}
 \int_0^{\infty} dx x^{2D-6}
 \nonumber
 \\
 & \times &
 \left[ - \frac{d}{dx} \left( \frac{ 1 + 2 x \cos \vartheta }{
 1 + 2 x \cos \vartheta + x^2} \right) \right]
 \;.
 \label{eq:Iprime}
 \end{eqnarray}
In Fig.~\ref{fig:BD} we show a numerical evaluation of $B_D$ for $3
< D < 4$.
With ${I}_3 = 1$ and $K_3^{\prime} = \pi^2/4$ we obtain for $D \rightarrow 3$
 \begin{equation}
 B_D \sim \frac{ b_3}{D-3}
 \; \; , \, \;
 b_3 = \frac{ 3 \pi^2 u_{\ast}^2}{16}
 \; .
 \label{eq:BDpole}
 \end{equation}
Perturbation theory is valid as long as the correction
(\ref{eq:selfpert2}) is small compared with $\rho_0 k^2$, i.~e.~for $
k \gtrsim k_c B_D^{1/2 \epsilon}$. Due to the IR divergence of $B_D$
for $D \rightarrow 3$, close to three dimensions perturbation theory
breaks down even outside the critical  regime. In a IR regularized
approach with a cutoff $\xi$, the divergence in
Eq.~({\ref{eq:selfpert2}) is replaced by a $\ln(k \xi)$ behavior
\cite{Baym01}. Below we shall see that the functional RG approach shows a
$\ln(k / k_c)$ behavior, as expected for a theory where 
the divergence due to density fluctuations is screened within a
non-perturbative treatment.

\subsection{Functional RG calculation of the self-energy}

We now derive the behavior of the self-energy microscopically using  the 
functional
RG equations for the effective classical field theory.

\subsubsection{The $u_l^2$-truncation for the $\dot{\Gamma}^{(4)}_l$
inhomogeneity}

To begin with, we need the RG flow of the four-point vertex.
In the simplest approximation, we
expand $\dot{\Gamma}_{l}^{(4) }
( {\bf{q}}_1^{\prime} , {\bf{q}}_2^{\prime} ; {\bf{q}}_2 , {\bf{q}}_1 ) $
in powers of $u_l$, which should be accurate
as long as the renormalized coupling remains small. To leading order we simply
ignore the six-point vertex and replace the four-point vertices
on the right-hand side of   Eq.~(\ref{eq:fourpointscale}) by their relevant part $u_l$.
 In this approximation,
  \begin{eqnarray}
 \dot{\Gamma}_{l}^{(4) }
( {\bf{q}}_1^{\prime} , {\bf{q}}_2^{\prime} ; {\bf{q}}_2 , {\bf{q}}_1 )
 & \approx &   - {u}_l^2
 \Bigl[ \frac{1}{2} \dot{\chi}_l ( |  {\bf{q}}_1 + {\bf{q}}_2 | )
 \nonumber
 \\
 & & \hspace{-20mm}
 + \dot{\chi}_l (| {\bf{q}}_1 - {\bf{q}}_{1}^{\prime}  | )
 + \dot{\chi}_l (| {\bf{q}}_1 - {\bf{q}}_{2}^{\prime} | )
 \Bigr]
 \; \;  ,
 \label{eq:dotGamma4approx}
 \end{eqnarray}

\noindent where the generalized susceptibility is given by

 \begin{eqnarray}
 \dot{\chi}_l ( q ) & = &  2 \int_{ {\bf{q}}^{\prime}}
 \dot{G} ( |  {\bf{q}}^{\prime} | )  \tilde{G}_l ( | {\bf{q}}^{\prime} + {\bf{q}}|  )
 \nonumber
 \\
 &   \approx &
\frac{2}{1+r_l} \Big<
\frac{
\Theta ( 1  < | {\bf{\hat{q}^\prime}} + {\bf{q}} |  < e^{  l} ) }{
  | {\bf{\hat{q}^\prime }} +  {\bf{q}} |^2  + r_l  } \Big>_{\bf \hat{q}^\prime}
 \nonumber
 \\
 &   = &
  \frac{2}{1+r_l} \frac{  \Omega_{D-1}}{ \Omega_D } \int_{0}^{\pi} d \vartheta
 ( \sin \vartheta )^{D-2}
 \nonumber
 \\
 & \times &
 \frac{
\Theta ( 0 <  q^2 + 2 q \cos \vartheta < e^{ 2 l} -1 ) }{
 1 + r_l + q^2  + 2 q \cos \vartheta  }
 \; ,
 \label{eq:dotchiapprox}
 \end{eqnarray}
where in the second and third  line we have used the approximations
(\ref{eq:Glapprox}) and (\ref{eq:dotGlapprox}),
taking into account only the relevant and marginal
part of the two-point vertex.

\subsubsection{Generalized susceptibility}

It is instructive to examine the behavior of the
generalized susceptibility $\dot{\chi}_l ( q )$.
For  $e^{l}-1 \gg 2$
the asymptotic behavior for small and large $q$ can be  easily obtained
analytically,
 \begin{equation}
 \dot{\chi}_l ( q ) \sim
 \left\{
 \begin{array}{ll}
 \frac{1}{ ( 1 + r_l )^2 }  + O (q )& \mbox{for $ q \ll 1 \ll e^{l}$} \, ,\\
 \frac{ 2 }{ ( 1 + r_l ) q^2 } & \mbox{for $ 1 \ll q \ll e^l$} \, .
 \end{array}
 \right.
 \label{chidotasym}
 \end{equation}
The important point is that the leading correction
for small $q$ is linear in $q$:
expanding Eq.~(\ref{eq:dotchiapprox})
in powers of $q$,
 \begin{equation}
 \dot{\chi}_l ( q ) = \dot{\chi}_l ( 0 ) +  \dot{\chi}_l^{\prime} ( 0 )  q
 + O (q^2 )
 \; ,
 \end{equation}
we find
 \begin{equation}
 \dot{\chi}_l ( 0 ) = \frac{1}{ ( 1 + r_l )^2 }
 \; ,
 \end{equation}
 \begin{equation}
 \dot{\chi}_l^{\prime} ( 0 ) = - \frac{1}{ ( 1 + r_l )^2 }
 \left[  \frac{4 S_1}{ ( 1 + r_l ) } - S_0
   \right]
 \; ,
 \end{equation}
where
 \begin{equation}
 S_0 =
 \big<
 \delta ( {\bf\hat{q}^\prime}
  \cdot \hat{\bf{q}} )
\big>_{\bf \hat{q}^\prime}
  = \frac{\Omega_{D-1}}{\Omega_D}
 \; ,
 \label{eq:s0def}
 \end{equation}
and
 \begin{equation}
 S_1 =
\big<
 \Theta ( {\bf \hat{q}^\prime} \cdot
 \hat{\bf{q}} )  {\bf \hat{q}^\prime} \cdot \hat{\bf{q}}
 \big>_{\bf \hat{q}^\prime}
 = \frac{S_0}{D-1}
 \; .
 \label{eq:s1def}
 \end{equation}
In $D=3$ we obtain with $S_0 = 1/2$ and $S_1 = 1/4$
\begin{equation}
 \dot{\chi}_l^{\prime} (0 ) = - \frac{1}{ (1 + r_l )^2 }
  \left( \frac{1}{ 1 + r_l } - \frac{1}{2} \right)
  \; \; , \; \; D=3
 \; .
 \label{eq:dotchiexp3}
 \end{equation}
For later reference, let us give the exact function  $\dot{\chi}_l ( q ) $
in $D=3$, where  the angular integration in Eq.~(\ref{eq:dotchiapprox}) can easily be
performed analytically.
The result can be written as
 \begin{equation}
 \dot{\chi}_l ( q ) = \frac{  \Theta ( q_2 - q_1 ) }{2 ( 1 + r_l ) q }
 \ln \left[ \frac{ 1 + r_l + q^2 + 2 q q_2}{
 1 + r_l + q^2 + 2 q q_1 } \right]
 \; ,
 \end{equation}
 where
 \begin{eqnarray}
 q_1 &=& \left\{ \begin{array}{ll}
 -1 & \hspace{.7cm} \mbox{if $ q > 2 $ } \\
 - q/2 & \hspace{.7cm} \mbox{if $ q < 2 $ }
 \end{array}
 \right.
 \label{eq:x1def}
 \; , \\
 q_2 &=& \left\{ \begin{array}{ll}
 1 & \mbox{if $  e^l -1  > q $ } \\
  \frac{ e^{2l} -1 }{2q} - \frac{q}{2}  & \mbox{if $ e^{l} -1  < q $ }
 \end{array}
 \right.
 \label{eq:x2def}
 \; .
 \end{eqnarray}
In particular, for  $q \leq {\rm min} \{ 2 , e^{l} - 1 \}$ we have
 \begin{equation}
 \dot{\chi}_l ( q ) = \frac{1}{2 ( 1 + r_l ) q }
 \ln \left[ \frac{ ( 1 + q )^2 + r_l }{
 1 + r_l  } \right]
 \label{eq:dotchi3small}
 \; .
 \end{equation}
Thus, for small momenta our approximation
(\ref{eq:dotGamma4approx}) yields in $D=3$,
  \begin{eqnarray}
 \dot{\Gamma}_{l}^{(4) }
( {\bf{q}}_1^{\prime} , {\bf{q}}_2^{\prime} ; {\bf{q}}_2 , {\bf{q}}_1 )
 & \approx &
 - \frac{5 }{2} \frac{ u_l^2 }{ ( 1 + r_l )^2 }
 \nonumber
 \\
 &  & \hspace{-30mm} +
 \frac{u_l^2}{ (1 + r_l )^2 }
  \left( \frac{1}{ 1 + r_l } - \frac{1}{2} \right)
 \Bigl[ \frac{1}{2} |  {\bf{q}}_1 + {\bf{q}}_2   |
 \nonumber
 \\
 &   & \hspace{-30mm}
 + | {\bf{q}}_1 - {\bf{q}}_{1}^{\prime} |
 +  |  {\bf{q}}_1 - {\bf{q}}_{2}^{\prime}  | \Bigr]
 + O ( {\bf{q}}_i^2 )
 \;  .
 \label{eq:dotGamma4approx2}
 \end{eqnarray}
The term linear in $q$ generates marginal parameters, as mentioned
in Sec.\ref{subsec:marginal}, when we iterate the RG. The linear
term exists for all $D$, but the corresponding coupling parameters
are irrelevant  in $D > 3$. Because   these coupling parameters are
not consistently  taken into account in the $u^2$-truncation given
in Eq.~(\ref{eq:dotGamma4approx}), we cannot expect that this
truncation gives numerically accurate results in the critical regime
close to three dimensions. On the other hand, in the
short-wavelength regime (and also for $ D - 3 \gg \eta$), our
$u^2$-truncation (\ref{eq:dotGamma4approx}) is sufficient. In this
section we shall therefore proceed with this approximation, which
produces well-defined results even in $D=3$ provided the anomalous
exponent $\eta$ is calculated self-consistently by solving an
integral equation, see Eq.~(\ref{eq:etaintegral}) below. In Sec.
\ref{sec:self3} we shall improve on this approximation by explicitly
including all marginal coupling parameters in the critical regime in
$D=3$.

\subsubsection{Explicit expression for the four-point vertex}

With Eq.~(\ref{eq:dotGamma4approx}) as an approximation for
$\dot{\Gamma}_l^{(4)}$, we are now in a position to calculate
the flow of the four-point vertex.
Let us rewrite Eq.~(\ref{eq:dotGamma4approx})
in the following way,
  \begin{eqnarray}
 \dot{\Gamma}_{l}^{(4) }
( {\bf{q}}_1^{\prime} , {\bf{q}}_2^{\prime} ; {\bf{q}}_2 , {\bf{q}}_1 )
  \approx
\frac{-5 u_l^2 }{2 ( 1 + r_l )^2 }
 + \dot{\Gamma}_{l}^{(4\rm mi) }
( {\bf{q}}_1^{\prime} , {\bf{q}}_2^{\prime} ; {\bf{q}}_2 , {\bf{q}}_1 )
 \;   , \nonumber \\ &&
 \label{eq:dotGamma4approx3}
 \end{eqnarray}
where
 \begin{eqnarray}
 \dot{\Gamma}_{l}^{(4\rm mi) }
( {\bf{q}}_1^{\prime} , {\bf{q}}_2^{\prime} ; {\bf{q}}_2 , {\bf{q}}_1 ) &=&
 - u_l^2 \Big[
\frac{1}{2} \dot{\chi}^{(\rm mi)}_l ( | {\bf{q}}_1 + {\bf{q}}_2  |)
\nonumber \\ && \left. \hspace{-3cm}
 + \dot{\chi}_l^{(\rm mi)} (| {\bf{q}}_1 - {\bf{q}}_{1}^{\prime} | )
 + \dot{\chi}_l^{(\rm mi)} (| {\bf{q}}_1 - {\bf{q}}_{2}^{\prime} | )
 \right]
 \; ,
 \label{eq:dotGamma4mi}
 \end{eqnarray}
with
\begin{equation}
 \dot{\chi}_l^{(\rm mi)} ( q ) = \dot{\chi}_l ( q )  -  \dot{\chi}_l (0)
 \label{eq:chiirrel}
 \;
 \end{equation}
describing the momentum-dependent part. The superscript $(\rm mi)$
indicates that these terms are marginal or irrelevant. In this
approximation the four-point vertex takes the form
 \begin{equation}
 \label{eq:Gamma4}
 \tilde{\Gamma}_{l}^{ (4)}
 ( {\bf{q}}_1^{\prime} , {\bf{q}}_2^{\prime} ; {\bf{q}}_2 , {\bf{q}}_1 )
 \approx u_l +  \tilde{\Gamma}_{l}^{(4\rm mi) }
( {\bf{q}}_1^{\prime} , {\bf{q}}_2^{\prime} ; {\bf{q}}_2 , {\bf{q}}_1 )
 \; ,
 \end{equation}
with the irrelevant (and in $D=3$ also marginal) parts  given by
\begin{widetext}
 \begin{eqnarray}
 \tilde{\Gamma}_{l}^{(4\rm mi) }
( {\bf{q}}_1^{\prime} , {\bf{q}}_2^{\prime} ; {\bf{q}}_2 , {\bf{q}}_1 )
 & = &
 \int_0^{l} d l^{\prime}
 e^{ \epsilon ( l - l^{\prime} ) - 2 \int_{l^{\prime}}^l d \tau \eta_{\tau} }
 \dot{\Gamma}_{l^{\prime}}^{(4\rm mi) }
( e^{ - ( l - l^{\prime} ) } {\bf{q}}_1^{\prime}  ,
e^{ - ( l - l^{\prime} ) }  {\bf{q}}_2^{\prime} ;   e^{ - ( l - l^{\prime} ) }  {\bf{q}}_2 ,
e^{ - ( l - l^{\prime} ) }  {\bf{q}}_1 )
 \nonumber
 \\
&  & \hspace{-30mm} = -
 \int_0^{l} d l^{\prime} e^{  \epsilon (l - l^{\prime})   -  2 \int_{l^{\prime}}^l d \tau \eta_{\tau}
 } u_{l^{\prime}}^2
 \left[ \frac{1}{2} \dot{\chi}^{(\rm mi)}_{l^{\prime}} (  e^{ - ( l - l^{\prime} )} | {\bf{q}}_1 + {\bf{q}}_2 | )
 + \dot{\chi}^{(\rm mi)}_{l^{\prime}} ( e^{ - ( l - l^{\prime} ) } | {\bf{q}}_1 - {\bf{q}}_1^{\prime} | )
 + \dot{\chi}^{(\rm mi)}_{l^{\prime}} ( e^{ - ( l - l^{\prime} ) } | {\bf{q}}_1 - {\bf{q}}_2^{\prime} | )
 \right]
 \nonumber
 \\
&  & \hspace{-30mm}  = -
 \int_0^{l} d t e^{  \epsilon t   -  2 \int_{l-t}^l d \tau \eta_{\tau}
 } u_{l - t}^2
 \left[ \frac{1}{2} \dot{\chi}^{(\rm mi)}_{l-t} (  e^{ - t} | {\bf{q}}_1 + {\bf{q}}_2 | )
 + \dot{\chi}^{(\rm mi)}_{l-t} ( e^{ - t } | {\bf{q}}_1 - {\bf{q}}_1^{\prime} | )
 + \dot{\chi}^{(\rm mi)}_{l-t} ( e^{ - t } | {\bf{q}}_1 - {\bf{q}}_2^{\prime} | )
 \right]
 \; .
 \label{eq:Gamma4final}
 \end{eqnarray}
\end{widetext}

\subsection{Anomalous dimension}

To obtain the anomalous dimension $\eta_l  = \partial
\dot{\Gamma}^{(2)}_l ( {\bf q} ) /
 \partial q^2  |_{ q^2 =0} $ we calculate
 $\dot{\Gamma}_l^{(2)} ( {\bf{q}} )$ via Eq.~(\ref{eq:gammadot2def}).
 With the approximation Eqs.~(\ref{eq:Gamma4}, \ref{eq:Gamma4final}) for the
four-point  vertex we find
  \begin{eqnarray}
 \dot{\Gamma}_l^{(2)} ( {\bf {q}} ) & = & \frac{1}{R_l ( 1 )}
\Big<
 \tilde{\Gamma}^{(4)}_l ( {\bf{q}} , {\bf{\hat{q}^\prime}} ;
 {\bf{\hat{q}^\prime}} , {\bf{q}} ) \Big>_{\bf \hat{q}^\prime}
 \nonumber
 \\
 & \approx & \frac{u_l}{1 + r_l }
 - \frac{3}{2(1+r_l)}
 \int_0^{l} d t e^{  \epsilon t   -  2 \int_{l-t}^l d \tau \eta_{\tau}
 } {u}_{l-t}^2
 \nonumber
 \\
 &  &  \hspace{10mm} \times \
 \big<
 \dot{\chi}^{(\rm mi)}_{l-t} (  e^{ - t} ( | {\bf{\hat{q}^\prime}} + {\bf{q}} |  )
 \big>_{\bf \hat{q}^\prime}
 \; .
 \label{eq:gammadot2approx}
 \end{eqnarray}
From Eq.~(\ref{eq:anomalexplicit}) we then obtain
 an  integral equation for the flowing anomalous dimension
 \begin{equation}
  \eta_l =    \int_0^{l} d t K (l,t)  u_{l-t}^2
 e^{  - 2 \int_{l-t}^l d \tau \eta_{\tau} }
 \; ,
 \label{eq:etaintegral}
 \end{equation}
with
 \begin{eqnarray}
 K ( l , t )   & = & -
 \frac{3 e^{\epsilon t} }{2 (1 + r_l)}
 \frac{\partial}{\partial q^2}
\Big<
 \dot{\chi}_{l-t} (  e^{ - t } ( |   {\bf{\hat{q}^\prime}} + {\bf{q}} |  ) )
 \Big>_{\bf \hat{q}^\prime}
 \Big|_{ q=0}
 \nonumber
 \\
 & = & - \frac{3}{4 D (1 + r_l)} \Bigl[ (D-1 ) e^{- ( D-3 ) t }
 \dot{\chi}^{\prime}_{l-t} ( e^{-t} )
 \nonumber
 \\
 & & \hspace{10mm} + e^{- ( D-2) t }
  \dot{\chi}^{\prime \prime}_{l-t} ( e^{-t} ) \Bigr]
 \; ,
 \label{eq:Kltdef}
 \end{eqnarray}
where $ \dot{\chi}^{\prime}_{l} ( x ) = d \dot{\chi}_l ( x ) / d x $ and
$ \dot{\chi}^{\prime \prime}_{l} ( x ) = d^2 \dot{\chi}_l ( x ) / d x^2 $.
Note that on the right-hand side we have replaced $\dot{\chi}^{(\rm mi)}_{l-t} (q )
\rightarrow \dot{\chi}_{l-t} ( q )$, because the constant part $\dot{\chi}_{l-t} ( 0 )$ does not contribute to the derivative.
The integral equation (\ref{eq:etaintegral})
together with the one-loop flow equations for the relevant coupling parameters,
 \begin{equation}
 \partial_l r_l = ( 2 - \eta_l ) r_l + \frac{u_l}{1 +r_l}
 \label{eq:rlflow2}
 \; ,
 \end{equation}
  \begin{equation}
 \partial_l u_l = ( \epsilon  - 2 \eta_l ) u_l  - \frac{5}{2} \frac{u_l^2}{(1 +r_l)^2}
 \label{eq:ulflow2}
 \; ,
 \end{equation}
form a system of three coupled integro-differential equations
for the three unknown functions $r_l$, $u_l$, and $\eta_l$.
A numerical solution of these equations for $D=3$ is shown
in Fig.~\ref{fig:etaflow}.
\begin{figure}
\epsfysize5.5cm
\hspace{5mm}
\epsfbox{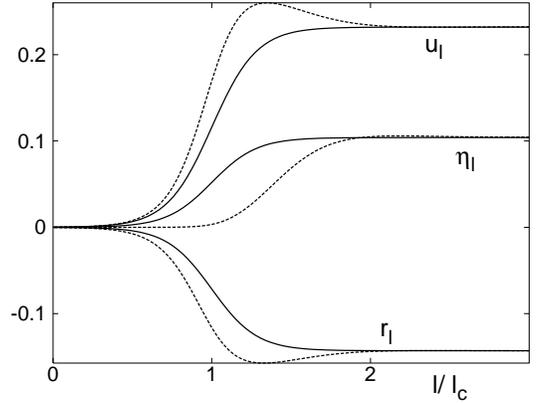} \vspace{5mm} \caption{  Self-consistent RG
flows of $u_l$, $r_l$ and $\eta_l$ at the critical point in  $D=3$
(dashed lines), obtained numerically from the coupled
intego-differential
Eqs.~(\ref{eq:etaintegral},\ref{eq:rlflow2},\ref{eq:ulflow2}). For
comparison, the Fermi-function approximation Eq.~(\ref{eq:etafermi})
for the flows is also shown (full lines), with fixed point values
matching the numerical solution. \label{fig:etaflow} }
\end{figure}

To leading order in $\epsilon$ (i.~e.~close to four dimensions)
Eqs.~(\ref{eq:etaintegral},\ref{eq:rlflow2},\ref{eq:ulflow2})
decouple
and one recovers the fixed point values $u_{\ast} = 2 \epsilon / 5$,
$r_{\ast} = - \epsilon / 5$ from Eq.~(\ref{WilsonFisherFP}).
To this order it is consistent to ignore $\eta_l$ and $r_l$ on the
right-hand side of Eq.~(\ref{eq:ulflow2}), so that it reduces to
Eq.~(\ref{eq:gflow}), with the
Fermi function solution for $u_l$ given in
Eq.~(\ref{eq:Fermi}). From the  numerical solution of
Eqs.~(\ref{eq:etaintegral}--\ref{eq:ulflow2}) for $D=3$  we find
that even if $r_l$ and $\eta_l$ are taken into account, to a good
approximation the qualitative behavior of  $u_l$ is still well
described by a Fermi function if we adjust the fixed point value
$u_*$ to the one obtained from the numerical solution. Furthermore,
Fig.~\ref{fig:etaflow} shows that also $\eta_l$ and $r_l$ roughly
follow the same functional form and we therefore approximate
 \begin{eqnarray}
 \frac{\eta_l}{\eta}   \approx \frac{r_l}{r_*} \approx \frac{ u_l}{ u_{\ast} }
\approx \frac{1}{e^{ \epsilon ( l_c - l ) } + 1 }
 \; ,
 \label{eq:etafermi}
 \end{eqnarray}
 with the fixed point values taken from the numerical solution
 and where $\eta = \lim_{l \rightarrow \infty} \eta_l$ is the anomalous
dimension at the fixed point. Although the functional form
(\ref{eq:etafermi}) for $\eta_l$ is qualitatively incorrect for $l
\ll l_c$ (where  $\eta_l \propto u_l^2$, whereas
Eq.~(\ref{eq:etafermi}) predicts $\eta_l \propto u_l$) we may use
Eq.~(\ref{eq:etafermi}) as a zeroth approximation on the right-hand
side of the integral equation (\ref{eq:etaintegral}). The point is
that for $l \ll l_c$ the right-hand side of
Eq.~(\ref{eq:etaintegral}) is not sensitive to the precise value of
$\eta_l$ in this regime, and the first iteration of the integral
equation yields an  accurate result even for small $l$. Let us now
calculate the fixed point values. We obtain from
Eq.~(\ref{eq:etaintegral})
 \begin{equation}
 \eta= u_{\ast}^2 \int_0^\infty d t K ( \infty , t ) e^{ - 2 \eta t }
 \; ,
 \label{eq:etaintegral2}
 \end{equation}
where we use the notation $\eta=\eta_*$.
The function $K ( \infty , t )$ is
obtained from $K ( l , t )$ defined in
Eq.~(\ref{eq:Kltdef})
by replacing the functions
$ \dot{\chi}_{l-t} ( q  )$ on the right-hand side
by
 \begin{eqnarray}
 \dot{\chi}_{\ast} ( q ) & \equiv & \lim_{ l \rightarrow \infty}
 \dot{\chi}_l ( q )
 \nonumber
 \\
 & \approx  &
 \frac{2}{1+r_*}
\Big<
 \frac{
\Theta (  | {\bf{\hat{q}^\prime }} + {\bf{q}} | -1 ) }{
 | {\bf{\hat{q}^\prime}} +  {\bf{q}} |^2  + r_\ast  }
\Big>_{\bf \hat{q}^\prime}
 \; .
 \label{eq:dotchiinftydef}
 \end{eqnarray}
In particular, in $D=3$ we have
 \begin{eqnarray}
 \dot{\chi}_{\ast} ( q ) & = & \frac{1}{2 ( 1 + r_{\ast} ) q }
 \Bigl[ \Theta ( 2 - q ) \ln \Bigl( \frac{ (1 + q )^2 + r_{\ast} }{ 1 +
 r_{\ast} } \Bigr)
 \nonumber
 \\
 & & \hspace{7mm} +  \Theta ( q - 2 ) \ln \Bigl( \frac{ (1 + q )^2 + r_{\ast} }{ ( 1-q )^2 +
 r_{\ast} } \Bigr)
 \Bigr]
 \; .
 \label{eq:dotchi3explicit}
 \end{eqnarray}
Note that for large $t$ the kernel
$K ( \infty , t )$ vanishes as $e^{ - ( D-3 ) t }$, so that
in $D=3$ it is crucial to retain the
$\eta$-dependence on the right-hand side of Eq.~(\ref{eq:etaintegral2}).
This is
closely related to the appearance of marginal terms of the four-point
vertex in $D=3$.
The self-consistent fixed point values
can be obtained by using the fact that, at the fixed point, we can express
$u_*$ and $r_*$ as a function of $\eta$,
\begin{subequations}
\begin{eqnarray}
  \label{FPofeta}
  r_*&=&- \frac{2}{3} \ \frac{1-2 \eta}{4-3\eta} \; ,
  \\
  u_*&=& \frac{10}{9}\
  \frac{(1-2 \eta)(2-\eta)^2}{(4-3\eta)^2} \; .
\end{eqnarray}
\end{subequations}
Inserting these expressions for $u_*$ and $r_*$ into
Eq.~(\ref{eq:etaintegral2}), we obtain a self-consistent equation
for $\eta$ which can be solved numerically. The numerical solution
gives
\begin{subequations}
 \begin{eqnarray}
 r_{\ast} & \approx &  -0.143
 \; ,
 \label{eq:rast}
 \\
 u_{\ast} &  \approx & 0.232
 \; ,
 \label{eq:uast}
 \\
 \eta & \approx  & 0.104
 \; .
 \label{eq:etaast1}
 \end{eqnarray}
\end{subequations}
The above value for $\eta$ is  approximately three times as large as
the  generally accepted value $\eta \approx  0.038$ obtained by
several different techniques \cite{Guida98,Campostrini01}. However,
given the simplicity of our truncation, it is quite satisfactory
that our estimate (\ref{eq:etaast1}) for $\eta$ has the correct
order of magnitude. In any case, our simple $u^2$-truncation
certainly gives a much better estimate for $\eta$ than the
self-consistent two-loop calculation employed by Baym et al.
\cite{Baym01}, which predicts a value of $0.5$ for $\eta$. In
Sec.\ref{sec:self3} we shall further improve on the $u^2$-truncation
employed here by explicitly taking the RG flow of marginal coupling
parameters into account.

For completeness, we show how to obtain the standard result for
$\eta$ to second order in an $\epsilon$-expansion. We can rewrite
Eq.~(\ref{eq:etaintegral2})
 as
 \begin{eqnarray}
\eta  &=&
 - \frac{3 u^2_{\ast}}{4 D ( 1 + r_{\ast})}
 \int_0^{1} d \lambda
\lambda^{ - \epsilon + 2 \eta }
 \Bigl[ ( D -1 )
 \dot{\chi}_{\ast}^{\prime} ( \lambda )
 +  \lambda
 \dot{\chi}_{\ast}^{\prime  \prime } ( \lambda )
 \Bigr]
 \nonumber
 \\
&=&    - \frac{3 u^2_{\ast}}{4 D ( 1 + r_{\ast})}
 \left[
2 (1 -  \eta ) \int_0^1 d \lambda \lambda^{ - \epsilon + 2 \eta }
 \dot{\chi}_{\ast}^{\prime} ( \lambda ) +  \dot{\chi}_{\ast}^{\prime} ( 1 )
 \right]
  . \nonumber \\
 \label{eq:etafixedapprox}
 \end{eqnarray}
For small $\epsilon = 4 - D$ we obtain to leading order
 \begin{equation}
   \label{epsilonexpand1}
 \eta = -\frac{3 u_{\ast}^2}{16}
 \left[
2  ( \dot{\chi} ( 1) -   \dot{\chi} ( 0 ))  + \dot{\chi}^{\prime} ( 1 )
 \right]
 \; ,
 \end{equation}
where the function $\dot{\chi} (q )$ is obtained from
$\dot{\chi}_{\ast} ( q )$ by simply setting $r_{\ast} \rightarrow
0$. Evaluating the integrals in $D=4$ (see also
Ref.~\cite{Kopietz01}),
\begin{subequations}
 \begin{eqnarray}
 \dot{\chi} ( 0 ) & = & 1
 \;,
 \\
 \dot{\chi} ( 1 ) & = & \frac{4}{3} - \frac{\sqrt{3}}{\pi}
 \; ,
\\
 \dot{\chi}^{\prime} ( 1 ) & = & - \frac{4}{3} + \frac{ 2 \sqrt{3}}{\pi}
 \; , \label{epsilonexpand4}
 \end{eqnarray}
\end{subequations}
and using $u_{\ast} = 2 \epsilon /5$, we finally obtain
$\eta = \epsilon^2 / 50 + O ( \epsilon^3)$, in agreement with
the field theoretical result \cite{ZinnJustin89}.

\subsection{The scaling function}

Let us now calculate the dimensionless scaling function
$\sigma_l ( x )$ which we define by
\begin{equation}
 \sigma_l ( x )  = e^{ - 2 ( l - l_c) } Z_l^{-1} \tilde{\Gamma}_l^{(2)}
  ( e^{ (l -l_c) } x )
 \; .
 \label{eq:sigmal2}
 \end{equation}
Since the two-point vertex $\tilde{\Gamma}_l^{(2)}({\bf q})$ and the
inhomogeneity $\dot{\Gamma}_l^{(2)}({\bf q})$ depend only on
$q=|{\bf q}|$, we shall in this subsection use scalar arguments for
these functions to simplify the notation. For $l \rightarrow \infty$
the definition (\ref{eq:sigmal2}) for $\sigma_l(x)$ yields the
universal scaling function defined in Eq.~(\ref{eq:sigmadef}),
 \begin{equation}
    \lim_{ l \rightarrow \infty} \sigma_l (x ) \equiv
 \sigma ( x ) = ( \rho_0 k_c^2 )^{-1} [ \Sigma ( k_c x ) - \Sigma (0 ) ]
 \; .
 \label{eq:sigmainfty}
 \end{equation}
We first perform some exact manipulations.
The functional flow equation  for  $\tilde{\Gamma}_l^{(2 )}  ( {{q}} )$
given by Eq.~(\ref{eq:flowtildeGamma2})
can be transformed into an integral equation
 \begin{eqnarray}
 \tilde{\Gamma}^{(2  )}_l ( {{q}} )
 & = & e^{ 2 l - \int_0^{l} d \tau \eta_{\tau} }
 \tilde{\Gamma}_{l=0}^{(2 )}  ( e^{-l} {{q}} )
 \nonumber \\ && +
\int_0^{l} d t e^{2 t - \int_{l-t}^{l} d \tau \eta_{\tau} }
 \dot{\Gamma}^{ (2)}_{ l - t } ( e^{-t} {{q}} )
 \nonumber
 \\
 & = & e^{ 2 l - \int_0^{l} d \tau \eta_{\tau} }
 \Bigg[ \tilde{\Gamma}_{l=0}^{(2 )}  ( e^{-l} {{q}} )  \nonumber \\
&& \hspace{-8mm}
\left. + \int_0^{l} dl^{\prime}
 e^{ - 2 l^{\prime} + \int_0^{l^{\prime}} d \tau \eta_{\tau} }
  \dot{\Gamma}^{ (2  )}_{ l^{\prime} } ( e^{-(l - l^{\prime})} {{q}} )
  \right] .
 \label{eq:gamma2int}
 \end{eqnarray}
To describe a critical system, we choose the initial value
 $\tilde{\Gamma}_{l=0}^{(2 )}  (  {{q}} ) = r_0$ to be momentum independent  such that for
$l \rightarrow \infty$ the
relevant coupling
 \begin{equation}
 r_l =  e^{ 2 l - \int_0^{l} d \tau \eta_{\tau} }
 \left[ r_0 +
 \int_0^{l} dl^{\prime}
 e^{ - 2 l^{\prime} + \int_0^{l^{\prime}} d \tau \eta_{\tau} }
  \dot{\Gamma}^{ (2  )}_{ l^{\prime} } ( 0 )
 \right]
\end{equation}
 has a finite limit $r_{\ast} = \lim_{ l \rightarrow \infty}
 r_l$, just has we have done in Eq.~(\ref{finetune1}).
This is guaranteed if the initial value $r_0$ is chosen such that
\cite{Ledowski03}
 \begin{equation}
 r_0  = -
 \int_0^{\infty} dl^{\prime}
 e^{ - 2 l^{\prime} + \int_0^{l^{\prime}} d \tau \eta_{\tau} }
  \dot{\Gamma}^{ (2  )}_{ l^{\prime} } ( 0 )
 \; .
 \label{eq:intial}
 \end{equation}
Defining
 \begin{eqnarray}
\dot{\Gamma}^{(2\rm mi)}_l ( q ) & = & \dot{\Gamma}^{(2)}_l ( q ) -  \dot{\Gamma}^{(2)}_l ( 0 )
 \label{eq:dotGamma2midef}
 \; ,
 \end{eqnarray}
where the superscript $mi$ indicates that these function contain only
{\it{marginal}} and {\it{irrelevant}} parameters,
we obtain
 \begin{eqnarray}
 \sigma_l ( x ) & = & e^{-2 ( l - l_c )} Z_l^{-1} r_l
 \nonumber
 \\
 &  & \hspace{-12mm} +
 \int_0^{l} dl^{\prime}
 e^{ - 2 ( l^{\prime} - l_c) + \int_0^{l^{\prime}} d \tau \eta_{\tau} }
  \dot{\Gamma}^{ (2 \rm mi )}_{ l^{\prime} } ( e^{ l^{\prime} - l_c } {{x}} )
 \; .
 \label{eq:sigmal3}
\end{eqnarray}
The first term on the right hand side vanishes for $l \rightarrow \infty$ because
by construction $\lim_{l \rightarrow \infty} r_l = r_{\ast}$ is finite on the critical surface.
Taking the limit $l \rightarrow \infty$ we thus obtain
(after renaming $l^{\prime} \rightarrow l$)
 \begin{equation}
 \sigma ( x ) =  \int_0^{\infty} dl
 e^{ - 2 ( l - l_c) + \int_0^{l} d \tau \eta_{\tau} }
  \dot{\Gamma}^{ (2 \rm mi )}_{ l } ( e^{ l - l_c } {{x}} )
 \; .
 \label{eq:sigmaexact}
 \end{equation}

\subsubsection{Truncation of the flow equations}

So far, no approximation has been made. We now approximate
the function $ \dot{\Gamma}^{ (2 \rm mi )}_{ l } ( q )$ on the right-hand side
 of Eq.~(\ref{eq:sigmaexact}) by the leading term in the expansion in powers
of $u_l$, see Eq.~(\ref{eq:gammadot2approx}),
  \begin{eqnarray}
 \dot{\Gamma}_l^{(2\rm mi)} ( {{q}} ) & \approx &
  - \frac{3}{2(1 + r_l)}
 \int_0^{l} d t e^{  \epsilon t   -  2 \int_{l-t}^l d \tau \eta_{\tau}
 } {u}_{l-t}^2
 \nonumber
 \\ & & \hspace{5mm} \times
 \big<
 \dot{\chi}^{}_{l-t} (  e^{ - t}  |  {\bf{\hat{q}^\prime}}+ {\bf{q}} |  )
 -  \dot{\chi}^{}_{l-t} (  e^{ - t}  )
 \big>_{\bf \hat{q}^\prime}
 \nonumber
 \\
 & & \hspace{-14mm} =   
- \frac{3}{2 ( 1 + r_l)}
 \int_0^{l} d l^{\prime} e^{  \epsilon ( l - l^{\prime})   -  2 \int_{l^{\prime}}^l d \tau \eta_{\tau}
 } {u}_{l^{\prime}}^2
 \nonumber
 \\ & & \hspace{-11mm} \times
 \big<
 \dot{\chi}^{}_{l^{\prime}} (  e^{ - ( l - l^{\prime})}  | {\bf{\hat{q}^\prime}}
+ {\bf{q}} |  )
 -  \dot{\chi}^{}_{l^{\prime}} (  e^{ - ( l - l^{\prime})}  )
 \big>_{\bf \hat{q}^\prime}
 \; .
 \label{eq:gamma2miapprox}
 \end{eqnarray}
All quantities on the right-hand side of Eq.~(\ref{eq:sigmaexact})
are now known, so that we may calculate the crossover function by
performing the four-dimensional integration (two angular
integrations and two integrations over the scale parameters $l$ and
$l^{\prime}$) numerically. There are no divergences even in $D=3$
provided the integral equation (\ref{eq:etaintegral}) for $\eta_l$
is solved self-consistently.

\subsubsection{Results for the scaling function}
\label{secscaling}
To obtain an analytic approximation for the
scaling function $\sigma ( x )$, we adopt again the
successful strategy used in the solution of the
integral equation (\ref{eq:etaintegral}) for $\eta$:
we substitute the Fermi function ansatz (\ref{eq:etafermi})
for the flowing anomalous dimension $\eta_{\tau}$
on the right-hand side of Eqs.~(\ref{eq:sigmaexact}) and
(\ref{eq:gamma2miapprox}).
After some  transformations of the integration variables
we then obtain
 \begin{equation}
 \sigma ( x ) = \frac{3 u_{\ast}^2}{2} x^{ 2 - \eta}
 \int_{ x e^{-l_c}  }^{\infty} d y
 \frac{ y^{ -3 + 2 \epsilon}}{[ x^{\epsilon} + y^{\epsilon} ]^{ 2 - 2 \eta / \epsilon} }
  F ( x, y; \eta , l_c)
 \; ,
 \label{eq:sigmares}
 \end{equation}
with
 \begin{eqnarray}
 F (  x , y ;\eta, l_c)
 & = & \int_{0}^{1} d z \frac{ z^{1 - \epsilon}}{ \left[ x^{\epsilon} + ( y/z )^{\epsilon}
 \right]^{ \eta / \epsilon} \bigl[
 1 + r_{l_c + \ln \frac{y}{zx} } \bigr] }
 \nonumber
 \\
 &  & \hspace{-22mm} \times
 \big<
 \dot{\chi}_{ l_c + \ln \frac{ y}{x} } ( z  )
 - \dot{\chi}_{ l_c + \ln \frac{y}{x} } ( | z {\bf{\hat{q}^\prime}}
+ y  {\bf{\hat{q}^{\prime \prime}}}  | )
 \big>_{\bf \hat{q}^\prime}
 \; ,
 \label{eq:FDdef}
 \end{eqnarray}
where ${\bf \hat{q}^{\prime\prime}}$ is an arbitrary unit vector.
Note that by assumption $x \ll e^{l_c}$
so that we may replace $x e^{-l_c} \rightarrow 0$
in the lower limit of the $y$-integral
in Eq.~(\ref{eq:sigmares}).
Consider first the regime $x \ll 1$.
Using the fact that for $l \gg l_c$ we may approximate
$r_l \rightarrow r_{\ast}$ and
$\dot{\chi}_l ( q ) \rightarrow
\lim_{ l \rightarrow \infty} \dot{\chi}_l ( q ) \equiv
\dot{\chi}_{\ast} (q ) $
(see Eq.~(\ref{eq:dotchiinftydef})), we may  replace in this regime
$ \dot{\chi}_{ l_c + \ln \frac{y}{x} } (q ) \rightarrow
 \dot{\chi}_{\ast} ( q )$ and
$r_{l_c + \ln \frac{y}{zx} } \rightarrow r_{\ast}$
in Eq.~(\ref{eq:FDdef}).
Then we obtain
 \begin{equation}
 \sigma ( x ) \approx  \frac{3 u_{\ast}^2}{2}  x^{ 2 - \eta}
 \int_{ 0  }^{\infty} d y
 \frac{ y^{ -3 + 2 \epsilon}}{[ x^{\epsilon} + y^{\epsilon} ]^{ 2 - 2 \eta / \epsilon} }
  F_{\ast} ( x, y; \eta )
 \; .
 \label{eq:sigmares2}
 \end{equation}
Using  $D$-dimensional spherical coordinates,  the function $ F_{\ast}
( x, y; \eta ) \equiv
 \lim_{l_c \rightarrow \infty}  F ( x, y; \eta,  l_c )$
can be written as
 \begin{eqnarray}
 F_{\ast} (  x , y ;\eta)
 & = &  \frac{\Omega_{D-1}}{(1 + r_{\ast}) \Omega_D}
  \int_{0}^{1} d z   \int_0^{\pi} d \vartheta     \frac{ z^{1 - \epsilon}   ( \sin \vartheta )^{ D-2}    }{ \left[ x^{\epsilon} + ( y/z )^{\epsilon}
 \right]^{ \eta / \epsilon} }
 \nonumber
 \\
 &    & \hspace{-10mm} \times
 \left[ \dot{\chi}_{\ast} ( z  )
 - \dot{\chi}_{\ast} ( \sqrt{ z^2 + 2 zy \cos \vartheta +  y^2 } )
 \right]
 \; .
 \label{eq:FDdef2}
 \end{eqnarray}
From Eqs.~(\ref{eq:sigmares2}) and (\ref{eq:FDdef2}) it is now straightforward
to obtain the asymptotic behavior of $\sigma (x )$ for small $x$,
 \begin{equation}
   \label{asymp:smallx}
 \sigma (x ) \sim A_D x^{ 2 - \eta }
 \; ,
 \end{equation}
 \begin{eqnarray}
 A_D & = &
\frac{3 u_{\ast}^2}{2}
 \int_{ 0  }^{\infty} d y y^{ -3 + 2 \eta}
  F_{\ast} ( 0 , y; \eta )
 \nonumber
 \\
 &  & \hspace{-14mm} =
\frac{3 u_{\ast}^2}{2(1 + r_{\ast})}
\frac{\Omega_{D-1}}{\Omega_D}
 \int_{ 0  }^{\infty} d y y^{ -3 +  \eta}
 \int_0^{1} dz   z^{1 - \epsilon + \eta}
\int_0^{\pi} d \vartheta
  \nonumber
 \\
 &     &  \hspace{-16mm} \times ( \sin \vartheta )^{ D-2}
 \left[ \dot{\chi}_{\ast} ( z  )
 - \dot{\chi}_{\ast} ( \sqrt{ z^2 + 2 zy \cos \vartheta +  y^2 } )
 \right]
  .
 \label{eq:AD}
 \end{eqnarray}
In $D=3$ we find numerically 
\begin{equation}
A_3 \approx 1.17.
\label{firstA3}
\end{equation}
On the other hand, for large $x$ the flow parameter in
$ \dot{\chi}_{ l_c + \ln \frac{y}{x} } (q ) $ and
$r_{l_c + \ln \frac{y}{zx} }$ is typically
small compared with $l_c$.
To take the effect of the flowing $r_l$
approximately into account, we replace in this regime
$r_{l_c + \ln \frac{y}{zx} } \rightarrow r_0$
and
$ \dot{\chi}_{ l_c + \ln \frac{y}{x} } (q )  \rightarrow
\dot{\chi}_0 ( q )$, where
$\dot{\chi}_0 (q )$ is obtained from
$\dot{\chi}_{\ast} (q )$ by replacing $r_{\ast} \rightarrow r_0$.
Using Eq.~(\ref{eq:r0fix}) one can show that on the critical surface
and for small $u_0$,
the initial value of $r_0$ is approximately given by
$r_0\approx-u_0/(2-\epsilon)$.
With this approximation we obtain
 in the regime $ [ 2 (D-3)]^{-1}  \ll \ln x$,
\begin{equation}
 \sigma (x ) \sim B_D x^{ 2  ( D-3) }
 \label{eq:sigmalargeasym}
 \; ,
 \end{equation}
 \begin{eqnarray}
 B_D & = &
\frac{3u_{\ast}^2}{2(1+r_0)}
\frac{\Omega_{D-1}}{\Omega_D}
 \int_{ 0  }^{\infty} d y y^{ -3 +  2 \epsilon }
 \int_0^{1} dz   z^{1 - \epsilon}
 \int_0^{\pi} d \vartheta
  \nonumber
 \\
 &     &  \hspace{-14mm} \times ( \sin \vartheta )^{ D-2}
 \left[ \dot{\chi}_0 ( z  )
 - \dot{\chi}_0 ( \sqrt{ z^2 + 2 zy \cos \vartheta +  y^2 } )
 \right]
 \; ,
 \label{eq:BD}
 \end{eqnarray}
and for $1 \ll \ln x \ll [ 2 (D-3)]^{-1}$ (which includes the limit $D \rightarrow 3$),
\begin{equation}
 \sigma (x ) \sim B_3^{\prime} \ln x + B_3^{\prime \prime}
 \label{asymp:largex}
 \; ,
 \end{equation}
where
 \begin{equation}
 B_3^{\prime} =
\frac{3u_{\ast}^2}{2(1+r_0)} \int_0^{1} dz \dot{\chi}_0 ( z )
 \approx \frac{ 3 \pi^2}{24} u_{\ast}^2
 \; \; ,
 \label{eq:Bprimeres}
\end{equation}
and on the right-hand side we have set $r_0 \rightarrow 0$.
Numerically we find $B_3^{\prime \prime} \approx 0.0319$. Note that
the physical self-energy
 \begin{equation}
 \Sigma ( k ) - \Sigma (0) = \rho_0 k_c^2 \sigma ( k / k_c )
 \label{eq:selfphysical}
\end{equation}
is independent of $u_{\ast}^2$ in the regime $k \gg k_c$: keeping in
mind that $k_c = \Lambda_0 ( u_0 / u_{\ast} )^{1/ \epsilon}$, it is
obvious that the factor of $u_{\ast}^2$ in Eqs.~(\ref{eq:BD}) and
(\ref{eq:Bprimeres}) combines with a suitable power of $k_c$ so that
the self-energy is proportional to $u_0^2$. A numerical evaluation
of $\sigma(x)$ is shown if Fig.~\ref{fig:sigma}.

\begin{figure}
\epsfxsize8.5cm
\epsfbox{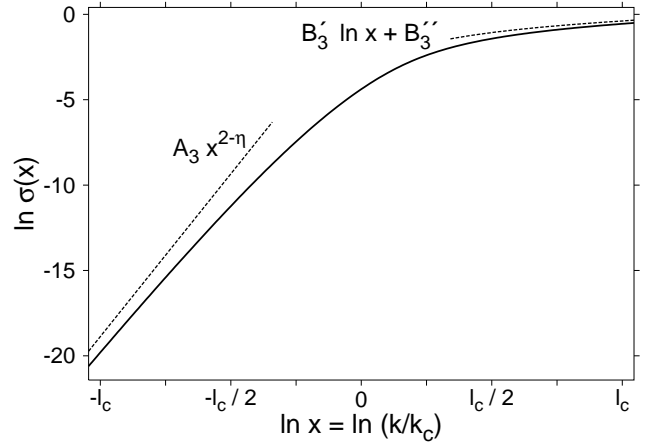}
\vspace{5mm}
\caption{
Numerical evaluation of the scaling function $\sigma(x)$ of the
self-energy in the limit $a\to 0$ and $D=3$.
Dashed lines are the
asymptotic forms predicted by Eqs.~(\ref{asymp:smallx},\ref{asymp:largex}).
}
\label{fig:sigma}
\end{figure}

To summarize, using the functional RG, we have derived in this section
the momentum-dependent  self-energy of weakly interacting
bosons, covering the entire range from the critical regime up to
momenta of the order of the inverse thermal de Broglie wavelength.
While in the critical regime $ k \ll k_c$ it is simpler to obtain
the asymptotic long-wavelength behavior of the self-energy within
the field theoretical RG, the functional RG approach adopted in this
work can describe the entire crossover from the critical regime $k
\ll k_c$ to the short-wavelength regime $k_c \ll k \lesssim 2 \pi /
\lambda_{\rm th} $. Note that the behavior of the self-energy in the
short-wavelength regime is determined by irrelevant terms which are
simply discarded in the field theoretical RG.

\section{Interaction induced $T_c$-shift}
\label{SECshift}

Baym {\em et al.} \cite{Baym01} have shown that to lowest order in
the scattering length $a$, the interaction induced shift for the
critical temperature $T_c$ in $D=3$ can be calculated from the ${\bf
k}$-dependence of the zero frequency self-energy $\Sigma({\bf
k},0)$. In a straightforward generalization of their result to
arbitrary dimensions, we can write the contribution from classical
modes to the shift in $T_c$ as
\begin{eqnarray}
 \frac{ \Delta T_c}{T_c}
 & = & \frac{2 \Omega_D}{\pi D \zeta ( D/2 )}
 \left( \frac{ k_c}{\Lambda_0 }\right)^{  D-2 }
 \nonumber
 \\
 &  & \hspace{-20mm} \times
 \int_0^{\Lambda_0 / k_c} d x x^{D-3}
 \frac{   ( \rho_0 k_c^2 )^{-1}  [ \Sigma ( k_c x ) - \Sigma (0)  ] }
{ x^2 +  ({\rho_0 k_c^2})^{-1}  [ \Sigma ( k_c x ) - \Sigma (0)  ] }
 \label{eq:tcshift}
 ,
 \end{eqnarray}
Using
$k_c / \Lambda_0 = e^{- l_c} = (u_0 / u_{\ast})^{1/ \epsilon }$, and
substituting our scaling function
$\sigma ( x)$ as defined in Eq.~(\ref{eq:sigmadef}), this can also be written as
 \begin{eqnarray}
 \frac{ \Delta T_c}{T_c}
 & = &
\frac{2 \Omega_D}{\pi D \zeta ( D/2 )}
 \left( \frac{ u_0}{u_{\ast} }\right)^{ \frac{ D-2}{4-D} }
 \nonumber
 \\ & &\times
 \int_0^{\Lambda_0 / k_c} d x x^{D-3}
 \frac{   \sigma ( x ) }
{ x^2 +  \sigma ( x ) }  .
 \label{eq:tcshift2}
 \end{eqnarray}
The shift in $T_c$ is dominated by classical fluctuations as long as
the UV cutoff $\Lambda_0 / k_c$ can be removed in
Eq.~(\ref{eq:tcshift2}). Keeping in mind that according to
Eq.~(\ref{eq:sigmalargeasym}) $\sigma ( x ) \sim B_D x^{ 2 ( D-3)}$
for large $x$, we see that the integral (\ref{eq:tcshift}) is UV
convergent as long as
 $ \frac{ D-2}{4-D} < 2$, i.~e.~$D < 10/3$. For
$D \geq  10/3$ the value of the integral in Eq.~(\ref{eq:tcshift})
depends on the UV cutoff, such that the lowest order contribution to
the shift in $T_c$ is  proportional to $ u_0^2$, with logarithmic
corrections ($ \propto u_0^2 \ln u_0$) in $D = 10/3$. Hence for $D
\geq  10/3$ the shift in the critical temperature cannot be obtained
from an effective classical field theory, while for $D<10/3$, the
cutoff only leads to higher order corrections (in $D=3$ one finds
$u_0^2$ and $u_0^2 \ln(u_0)$ corrections \cite{Baym01}). Writing
\begin{eqnarray}
 \frac{ \Delta T_c}{T_c} = J_D  u_0^{  \frac{ D-2}{4-D} }
 \label{eq:JDdef}
 \; ,
 \end{eqnarray}
we obtain for  $D < 10/3$,
 \begin{equation}
 J_D = \frac{2 \Omega_D}{\pi D \zeta ( D/2 )}
 \left( \frac{1}{u_{\ast} }\right)^{ \frac{ D-2}{4-D} }
 \int_0^{\infty} d x x^{D-3}
 \frac{   \sigma ( x ) }
{ x^2 +  \sigma ( x ) }
 \label{eq:JD}
 \; .
 \end{equation}
Note that $J_D$ implicitly depends on $u_0$ via $\sigma ( x )$.
However, for $u_0 \rightarrow 0$ the coefficient $J_D$
approaches a finite limit independent of $u_0$
as  long as  $D < 10/3$.

In $D = 3$ one usually writes
\begin{eqnarray}
  \frac{ \Delta T_c}{T_c}  &=& c_1 a n^{1/3}.
\end{eqnarray}
Keeping in mind that
 \begin{equation}
 u_0 = 16 \pi^{-1} [ \zeta ( 3/2 ) ]^{-1/3} a n^{1/3} \; \; \mbox{for $D=3$}
 \; ,
 \end{equation}
 we have
in this case $c_1 = 16 \pi^{-1} [ \zeta ( 3/2 ) ]^{-1/3} J_3$.
Within our approximation we find $c_1 = 1.23$, in rather good
agreement with the most accurate numerical investigations which give
$c_1=1.30 \pm0.02$ \cite{MC01a} and $c_1=1.29 \pm 0.05$
\cite{MC01b}. Variational perturbation techniques give similar
results, $c_1=1.23 \pm 0.12$ \cite{Kleinert03}, and $c_1=1.27 \pm
0.11$ \cite{Kastening03}. More results from other approaches are
summarized in a recent review \cite{Andersen04}. There is also
experimental data from a $^4$He-Vycor system confirming the linear
scaling of $T_c$ with $a n^{1/3}$ \cite{Reppy00}. The reported value
$c_1\approx 4.66$ seems however rather large which may be attributed
to uncertainties regarding the precise value of $a$ in the
experiment, see the discussion in Ref.~\cite{MC01b}.

\section{Improved description of the critical regime in $D=3$:
including marginal terms}
\label{sec:self3}

Within the $u^2$-truncation employed so far, we have neglected
the six-point vertex and
approximated the inhomogeneity 
in the flow equation
of the
four-point vertex 
by
  \begin{eqnarray}
 \dot{\Gamma}_{l}^{(4) }
( {\bf{q}}_1^{\prime} , {\bf{q}}_2^{\prime} ; {\bf{q}}_2 , {\bf{q}}_1 )
 & \approx &   - {u}_l^2
 \Bigl[ \frac{1}{2} \dot{\chi}_l ( |  {\bf{q}}_1 + {\bf{q}}_2 | )
 \nonumber
 \\
 & & \hspace{-20mm}
 + \dot{\chi}_l (| {\bf{q}}_1 - {\bf{q}}_{1}^{\prime}  | )
 + \dot{\chi}_l (| {\bf{q}}_1 - {\bf{q}}_{2}^{\prime} | )
 \Bigr]
 \; \;  ,
 \label{eq:dotGamma4approx3again}
 \end{eqnarray}
see Eq.~(\ref{eq:dotGamma4approx}).
This approximation amounts to a truncation of the exact
hierarchy of flow equations where
on the right-hand side of  Eq.~(\ref{eq:fourpointscale})
we simply replace

\begin{eqnarray}
 \tilde{\Gamma}_{l}^{ (4)}
 ( {\bf{q}}_1^{\prime} , {\bf{q}}_2^{\prime} ; {\bf{q}}_2 , {\bf{q}}_1 )
 & \rightarrow & u_l
 \; .
 \label{eq:fourpointapprox0}
 \end{eqnarray}
A formal justification for this procedure can only be given for $D>
3$: for weak interactions, irrelevant coupling parameters can be
expanded in powers of relevant and marginal ones
\cite{Polchinski84,Kopietz01}. For $3 < D < 4$ the only part
of the four-point vertex which is not irrelevant is its
constant part $u_l$. However, in $D=3$ there are  two additional
marginal parameters related to the momentum dependence of the four-point
vertex. Hence, in this case Eq.~(\ref{eq:dotGamma4approx3again}) is
not consistent. For small momenta the flowing four-point vertex has
the form
 \begin{eqnarray}
 \tilde{\Gamma}_{l}^{ (4)}
 ( {\bf{q}}_1^{\prime} , {\bf{q}}_2^{\prime} ; {\bf{q}}_2 , {\bf{q}}_1 )
 & = & u_l
 \nonumber
 \\
 &  & \hspace{-15mm} + {a}_l
 \left( | {\bf{q}}_1 - {\bf{q}}_1^{\prime} |+  | {\bf{q}}_1 - {\bf{q}}_2^{\prime} |
 \right) + b_l | {\bf{q}}_1 + {\bf{q}}_2 |
 \nonumber
 \\
 &  & \hspace{-15mm} + \tilde{\Gamma}_{l}^{ (4 \rm i)}
 ( {\bf{q}}_1^{\prime} , {\bf{q}}_2^{\prime} ; {\bf{q}}_2 , {\bf{q}}_1 )
 \label{eq:fourpointexp}
 \; ,
 \end{eqnarray}
where the irrelevant part  $\tilde{\Gamma}_{l}^{ (4 \rm i)}
 ( {\bf{q}}_1^{\prime} , {\bf{q}}_2^{\prime} ; {\bf{q}}_2 , {\bf{q}}_1 )$
 vanishes at least quadratically if all momenta become small.
 Below, we will denote the diagrams entering the renormalization of the
 four-point vertex using the usual fermionic language (i.~e.
BCS, ZS and ZS$^\prime$, see Fig.~\ref{fig:FourPoint}). 
We emphasize that this terminology simply
describes the topology of the diagrams, not the physical phenomena
arising from the diagrams.
Note that the four-point vertex must be symmetric under the exchange
${\bf{q}}_1 \leftrightarrow {\bf{q}}_2$ and ${\bf{q}}_1^{\prime}
\leftrightarrow {\bf{q}}_2^{\prime}$, so that both ``zero sound
channels'' are characterized  by the same parameter $a_l$. The
parameter $b_l$ characterizes the BCS channel.

 The exact flow
equations for the coupling parameters ${a}_l $ and ${b}_l$ are
 \begin{eqnarray}
 \partial_l {a}_l  & = &  ( 3 - D - 2 \eta_l ) {a}_l + \dot{\Gamma}_l^{a}
 \; ,
 \label{eq:flowa}
 \\
\partial_l {b}_l  & = &  ( 3 - D  - 2 \eta_l ) {b}_l + \dot{\Gamma}_l^{b}
 \; ,
 \label{eq:flowb}
 \end{eqnarray}
where  the coefficients $\dot{\Gamma}_l^{a}$ and
$\dot{\Gamma}_l^{b}$ are defined via the expansion of the function
$\dot{\Gamma}_{l}^{(4) } ( {\bf{q}}_1^{\prime} , {\bf{q}}_2^{\prime}
; {\bf{q}}_2 , {\bf{q}}_1 )$ in Eq.~(\ref{eq:fourpointscale}) for
small momenta,
 \begin{eqnarray}
 \dot{\Gamma}_{l}^{(4) }
( {\bf{q}}_1^{\prime} , {\bf{q}}_2^{\prime} ; {\bf{q}}_2 ,
{\bf{q}}_1 )
 & = &  \dot{\Gamma}_{l}^{(4) } (0,0; 0,0)
 \nonumber
 \\
 & + &  \dot{\Gamma}_l^{a}
 \left( | {\bf{q}}_1 - {\bf{q}}_1^{\prime} | + | {\bf{q}}_1 - {\bf{q}}_2^{\prime} | \right)
 \nonumber
 \\
 & +&    \dot{\Gamma}^{b}_l | {\bf{q}}_1 + {\bf{q}}_2 |
 + O ( {\bf{q}}_i^2 )
 \; .
 \label{eq:dotGammaabcdef}
 \end{eqnarray}
Even if the initial vertex
 $\tilde{\Gamma}_{l=0}^{ (4)}
 ( {\bf{q}}_1^{\prime} , {\bf{q}}_2^{\prime} ; {\bf{q}}_2 , {\bf{q}}_1 ) $
 is momentum-independent (corresponding to the initial conditions
${a}_0 = {b}_0 =0$), finite values of these parameters are generated
as we iterate the RG. Because in $D=3$ these  are marginal, they
cannot be ignored and are possibly the source of logarithmic
corrections.

Besides $a_l$ and $b_l$, there is a third   marginal parameter
in $D=3$, the
momentum-independent part of the six-point vertex,
\begin{eqnarray}
 \hspace{-20mm} {v}_l = \tilde{\Gamma}^{(6)}_l (0,0,0 ; 0,0,0 )
 \; ,
 \label{eq:vldef}
 \end{eqnarray}
see Eq.~(\ref{eq:rescaledflowgeneral}).
It satisfies the exact flow equation
 \begin{equation}
 \partial_l {v}_l
  =
 (  6 - 2 D - 3 \eta_l ) {v}_l
 +  \dot{\Gamma}_{l}^{(6) }
( 0,0,0 ; 0,0,0 )
 \;  .
 \label{eq:exactflowvl}
 \end{equation}
Again, the RG flow generates a finite ${v}_l$ even if initially
${v}_0 =0$. At the Wilson-Fisher fixed point in $D=3$ the
renormalized ${v}_{\ast} = \lim_{l \rightarrow \infty} {v}_l$ is
of  order  unity.

To arrive at the RG equations, we adopt the following approximation
scheme:
We expand $\dot{\Gamma}_l^{(n)}$ to
some fixed power in the marginal parameters
$a_l, b_l, v_l,$ and the relevant parameter $u_l$ while keeping higher
orders in the relevant parameter $r_l$.
The truncation is based on the hope that, although $a_l$, $b_l$, $v_l$ and
$u_l$ are not parametrically small, these parameters still remain
numerically small. The fixed point values for these parameters
which we obtain within
this scheme in Sec.~\ref{FPsection} are indeed much smaller than
unity, which gives some a posteriori justification to our approach.
To implement the scheme, it is useful to
transform Eq.~(\ref{eq:rescaledflowgeneral}) into an integral
equation,
\begin{eqnarray}
 \tilde{\Gamma}^{(n)}_l ( {\{ {\bf q}_i \}} )
 & = & e^{ [(2-D)n+D] l - n \int_0^{l} d \tau \eta_{\tau} }
 \tilde{\Gamma}_{l=0}^{(n)}  ( \{ e^{-l} { {\bf q}_i \}} )
  \nonumber \\ &&
\hspace{-2.cm} +
\int_0^{l} d t e^{[(2-D)n+D] t - n \int_{l-t}^{l} d \tau \eta_{\tau} }
 \dot{\Gamma}^{ (n)}_{ l - t } ( \{ e^{-t} {\bf q}_i \} )
\; .
 \label{eq:gammaNint}
 \end{eqnarray}
We use these retarded expressions in the calculations of the 
inhomogeneities
$\dot{\Gamma}_l^{(m)}$. Since the 
right hand side in Eq.~(\ref{eq:gammaNint}) depends
on vertices which also have the form (\ref{eq:gammaNint}), one quickly
arrives at rather complex expressions which we need to truncate. 
We will only include terms up to second
order in  $a_l, b_l, v_l,$ and $u_l$. 
However, even with this approximation
the calculation becomes rather involved.
To keep
the calculations tractable, we truncate further:  
To evaluate the flow of parameters associated with the $n$-point
vertex, 
we ignore irrelevant contributions arising from
vertices of order $m$
with $m>n+2$. Thus, when we calculate the flow of parameters associated
with the $n$-point vertex and
encounter a vertex $\tilde{\Gamma}_l^{(m)}$
in the calculation with $m>n+2$, 
we only keep its relevant and marginal
part instead of employing Eq.~(\ref{eq:gammaNint}).
Effectively, we neglect in this way infinitely many irrelevant
parameters related to higher order vertices. The expectation is,
that these irrelevant contributions are numerically small
compared to the contribtutions  originating from irrelevant terms of
lower order vertices.
In concrete terms, our truncation is as follows: When we calculate the
flow of $v_l$, we will consider irrelevant contributions from
$\tilde{\Gamma}_l^{(8)}$ but ignore contributions from
$\tilde{\Gamma}_l^{(10)}$ (in this case, there are in fact no
second order terms arising from inhomogeneities $\dot{\Gamma}_l^{(n)}$
with $n\ge 10$ so that we really have all second order terms
included in our flow of $v_l$). Similarly, to calculate the
flow of the four-point vertex parameters $a_l$, $b_l$, and $u_l$,
we ignore contributions arising from $\tilde{\Gamma}_l^{(8)}$. This
leads to the absence of $v_l^2$ terms in the RG equations for
$a_l$, $b_l$, and $u_l$, the
$\tilde{\Gamma}_l^{(8)}$ terms only enter indirectly via the flow
of $v_l$. To calculate the flow of $r_l$ we ignore irrelevant
contributions from $\tilde{\Gamma}_l^{(n)}$ for $n\ge 6$ which
again only enter via their contributions to the flow
of $a_l$, $b_l$, and $u_l$.

\subsection{Truncating the flow equation for the six-point
vertex}
We begin with the renormalization of $v_l$, the momentum independent
part of $\tilde{\Gamma}^{(6)}_l$, using the flow equations for the
six-point vertex from Appendix \ref{AppMarg},
Eq.~(\ref{eq:sixpoint}). Note that, working consistently  to second
order, the last two terms in Eq.~(\ref{eq:sixpoint}), which are at
least of third order in the marginal and relevant parameters, should
not be included. However, there is a contribution proportional to
$v_l^2$ arising from the eight-point vertex which must be included.
To see this, let us explicitly write $\dot{\Gamma}_l^{(6)}$ up to
second order in the marginal and relevant parameters (and zeroth
order in $\eta$),
\begin{widetext}
\begin{eqnarray}
\dot{\Gamma}_l^{(6)}( {\bf{q}}_1^\prime ,{\bf{q}}_2^\prime  ,
{\bf{q}}_3^\prime ; {\bf{q}}_3 , {\bf{q}}_2 , {\bf{q}}_1 ) &=&
\frac{-3 v_l}{1+r_l}\left\{3 {\cal S}_{(1,2),3}\left< G_l({\bf
q}_1+{\bf q}_2-{\bf \hat{q}}) (u_l+a_l(|{\bf q}_1-{\bf \hat{q}}| +
|{\bf q}_2-{\bf \hat{q}}|) +b_l |{\bf q}_1+{\bf q}_2|) \right>_{{\bf
\hat{q}}} \right. \nonumber
\nonumber \\ && \hspace{-10mm} +3\, {\cal
S}_{(1^\prime,2^\prime),3^\prime} {\cal S}_{(1,2),3} \left< G_l({\bf
q}_3-{\bf q}_3^\prime+{\bf \hat{q}}) (u_l+a_l(|{\bf q}_3-{\bf
q}_3^\prime |)+b_l| {\bf q}_3+{\bf \hat{q}}|) \right>_{{\bf \hat{q}}}
+({\bf q}_i \leftrightarrow {\bf q}_i^\prime ) \Big\}
\nonumber \\
&& \hspace{-10mm}+ \frac{1}{1+r_l}\int_0^l dt \ e^{-(4D-8)t} \left<
\dot{\Gamma}^{(8)}_{l-t} (e^{-t}{\bf q}_1^\prime,\dots,e^{-t}{\bf
q}_3^\prime,e^{-t}{\bf \hat{q}}; e^{-t}{\bf \hat{q}}, e^{-t}{\bf
q}_3,\dots,e^{-t}{\bf q}_1) \right>_{{\bf \hat{q}}} \; ,
\label{Gamma6punkt}
\end{eqnarray}
\end{widetext}
where we used
\begin{eqnarray}
\tilde{\Gamma}^{(8)}_l( {\bf{q}}_1^\prime ,\dots,{\bf{q}}_4^\prime ;
{\bf{q}}_4 ,\dots ,{\bf{q}}_1 ) &=& \int_0^l dt \ e^{-4 \int_{l-t}^l
d\tau \eta_\tau+(8-3D)t}
 \nonumber \\
 && \hspace{-4cm} \times \
\dot{\Gamma}^{(8)}_{l-t}(e^{-t} {\bf{q}}_1^\prime , \dots , e^{-t}
{\bf{q}}_4^\prime ;e^{-t}  {\bf{q}}_4 ,\dots , e^{-t} {\bf{q}}_1 )
\; , \label{eightpointscale}
\end{eqnarray}
and from Eq.~(\ref{eq:eightpoint}) we have
\begin{eqnarray}
\hspace{-1cm}
\dot{\Gamma}^{(8)}_l({\bf{q}}_1^\prime ,\dots,{\bf{q}}_4^\prime
; {\bf{q}}_4 ,\dots ,{\bf{q}}_1 )
&=&
\nonumber \\
&& \hspace{-4.6cm}
- 2v_l^2 \bigg\{
16\, {\cal S}_{(1,2,3),4}{\cal S}_{(1^\prime,2^\prime,3^\prime),4^\prime}
\, \dot{\chi}_l({\bf{q}}_1^\prime+{\bf{q}}_2^\prime+{\bf{q}}_3^\prime
-{\bf{q}}_4) 
\nonumber \\ 
&& \hspace{-4.6cm}
+18 \, {\cal S}_{(1,2),(3,4)}{\cal S}_{(1^\prime,2^\prime),(3^\prime,4^\prime)}
\, \dot{\chi}_l({\bf{q}}_1+{\bf{q}}_2-{\bf{q}}_1^\prime-{\bf{q}}_2^\prime)
\bigg\}
\nonumber \\ 
&& \hspace{-4.6cm}
+\ \mbox{terms at least cubic in $v_l$, $a_l$, $b_l$, and $u_l$}
\; . \label{Gamma8punkt}
\end{eqnarray}
The definition of the symmetrization operators ${\cal S}_{(1,2),3}$
used in Eq.~(\ref{Gamma6punkt}) and similar ones used in
Eq.~(\ref{Gamma8punkt}) can be found in
Eqs.~(\ref{A123b},\ref{A1234a},\ref{A1234b}) of Appendix \ref{AppMarg}.
 Keeping only the zeroth order in a momentum expansion, we
have
\begin{eqnarray}
&&\dot{\Gamma}^{(6)}_l(\{ {\bf q}_i\equiv 0\})
  =  -    12 \beta_0 v_l u_l
\label{eq:sixpointapprox2}
 -  15 \beta_0  v_l a_l
 -   9 \beta_0  v_l b_l
\\ && - 
\nonumber
\frac{2}{1+r_l} \int_0^l dt \ e^{-(4D-8)t}  v_{l-t}^2
\big[16 \dot{\chi}_{l-t}(0)+18\dot{\chi}_{l-t}(e^{-t})
\big] \; ,
\end{eqnarray}
where
\begin{eqnarray}
 \beta_0 & = & \dot{\chi}_l ( 0 )  = \frac{1}{ (1 + r_l )^2 }
 \; .
\label{eq:alpha0def}
\end{eqnarray}

\subsection{Improved truncation of the flow equation for the four-point
vertex}

Including the terms which become marginal in $D=3$ we need to
calculate

\begin{eqnarray}
 \dot{\Gamma}_{l}^{(4) }
( {\bf{q}}_1^{\prime} , {\bf{q}}_2^{\prime} ; {\bf{q}}_2 , {\bf{q}}_1 )
 & \approx  &
\frac{1}{1+r_l}
\big<
 \tilde{\Gamma}^{(6)}_l ( {\bf{q}}_1^\prime ,
{\bf{q}}_2^\prime , {\bf{\hat{q}}} ; {\bf{\hat{q}}} ,
 {\bf{q}}_2 , {\bf{q}}_1 ) \big>_{\bf \hat{q}}
\nonumber
 \\
 & + &
  \dot{\Gamma}_{l}^{(4,{\rm BCS}) }
( {\bf{q}}_1^{\prime} , {\bf{q}}_2^{\prime} ; {\bf{q}}_2 , {\bf{q}}_1 )
\nonumber
 \\
 & +  &
 \dot{\Gamma}_{l}^{(4,{\rm ZS}) }
( {\bf{q}}_1^{\prime} , {\bf{q}}_2^{\prime} ; {\bf{q}}_2 , {\bf{q}}_1 )
 \nonumber
 \\
 & + &
 \dot{\Gamma}_{l}^{(4,{\rm ZS}^{\prime}) }
( {\bf{q}}_1^{\prime} , {\bf{q}}_2^{\prime} ; {\bf{q}}_2 , {\bf{q}}_1 )
 \; ,
 \label{eq:dotGammaabcagain}
 \end{eqnarray}

\noindent where, instead of
Eqs.~(\ref{eq:dotGamma4approx3}, \ref{eq:dotGamma4mi}), we
now have to calculate the zeroth and first order terms in a momentum
expansion
of  Eq.~(\ref{eq:dotGammaabcagain}) in powers of both
the relevant and marginal parameters.
We discuss the terms in Eq.~(\ref{eq:dotGammaabcagain})
below, beginning with the contribution from
the six-point vertex.

\subsubsection{Contributions from $\tilde{\Gamma}_l^{(6)}$}

To calculate
$\big< \tilde{\Gamma}^{(6)}_l ( {\bf{q}}_1^\prime , {\bf{q}}_2^\prime ,
{\bf{\hat{q} }} ; {\bf{\hat{q} }} ,
 {\bf{q}}_2 , {\bf{q}}_1 ) \big>_{\bf \hat{q}}$,
we write, assuming $\tilde{\Gamma}^{(6)}_{l=0}=0$,

\begin{eqnarray}
\tilde{\Gamma}^{(6)}_l( {\bf{q}}_1^\prime , {\bf{q}}_2^\prime ,
{\bf{\hat{q}}} ; {\bf{\hat{q}}} , {\bf{q}}_2 , {\bf{q}}_1 ) &=&
\int_0^l dt\ e^{-3 \int_{l-t}^l d\tau \eta_\tau+(6-2D)t}
\nonumber
\\ &&  \hspace{-4.2cm} \times \ \dot{\Gamma}^{(6)}_{l-t}(e^{-t}
{\bf{q}}_1^\prime , e^{-t} {\bf{q}}_2^\prime , e^{-t} {\bf{\hat{q}}}
;e^{-t}  {\bf \hat{q}} , e^{-t}  {\bf{q}}_2 , e^{-t} {\bf{q}}_1 ).
\label{sixpointscale}
\end{eqnarray}
Using Eqs.~(\ref{sixpointscale},\ref{Gamma6punkt},\ref{Gamma8punkt}),
we can calculate the contribution of $\tilde{\Gamma}^{(6)}_l$ to
$\dot{\Gamma}^{(4)}_l$.
Including from the eight-point vertex only the 
${\bf q}_i=0$ contribution and
expanding the other contributions up to linear 
order in the external momenta we have,
\begin{widetext}
\begin{eqnarray}
\label{DotSixPoint2}
\Big< \dot{\Gamma}_l^{(6)}(\lambda {\bf q}_1^\prime,\lambda{\bf q}_2^\prime,
\lambda {\bf \hat{q}}, \lambda
{\bf \hat{q}}, \lambda {\bf q}_2, \lambda {\bf q}_1) \Big>_{\bf\hat{q}}
&\approx&
-\frac{v_l}{2} \Big\{12 u_l (\beta_0+\dot{\chi}_l(\lambda))\nonumber
+a_l\left( 12\beta_0+8  \dot{\chi}_l(\lambda)(1+\lambda)+8\dot{\Phi}_l(\lambda)
+2 \dot{\varphi}_l(\lambda)
\right)
\\ && \hspace{-2.2cm} \nonumber
+b_l\left( 8\beta_0+4  \dot{\chi}_l(\lambda)(1+\lambda)+4 \dot{\Phi}_l(\lambda)
+2 \dot{\varphi}_l(\lambda)
\right)
+\lambda |{\bf q}_1+{\bf q}_2| \left(2 \beta_1 u_l +(4\beta_1+2 \beta_2)a_l+2\beta_0 b_l
\right)
\\ && \hspace{-2.2cm}
+\lambda \left(|{\bf q}_1-{\bf q}_1^\prime|+|{\bf q}_1-{\bf q}_2^\prime|\right)
\left( 4\beta_1 u_l+(4\beta_0+4\beta_1+2 \beta_2)a_l+(4\beta_1+2 \beta_2)b_l\right)
\Big\}
\nonumber \\
&& \hspace{-2.2cm}
-\frac{2}{1+r_l} \int_0^l dt \ e^{-(4D-8)t}  v_{l-t}^2
\big[16 \dot{\chi}_{l-t}(0)+18\dot{\chi}_{l-t}(e^{-t})
\big] \; ,
\end{eqnarray}
\end{widetext}
where
\begin{eqnarray}
 \beta_1 & = & \dot{\chi}^{\prime}_l ( 0 ) =
  \frac{ S_0}{ ( 1 + r_l )^2}  - \frac{4 S_1}{ ( 1 + r_l )^3 }
 \label{eq:alpha1def}
 \; ,
 \\
 \beta_2 & = &
 \frac{2 S_1}{ ( 1 + r_l )^2 }
 \label{eq:alpha2def}
 \; ,
\end{eqnarray}
and $S_0$ and $S_1$ are defined in Eqs.~(\ref{eq:s0def}) and
(\ref{eq:s1def}). We further introduced the functions

\begin{eqnarray}
\dot{\Phi}_l(\lambda)=\frac{2}{1+r_l} \Big< \tilde{G}_l({\bf \hat{q}}+{\bf \hat{q}^{\prime}}\lambda)
|{\bf \hat{q}}+{\bf \hat{q}^{\prime}}\lambda| \Big>_{\bf \hat{q}} \; ,
\end{eqnarray}
and

\begin{eqnarray}
\dot{\varphi}_l(\lambda)=\frac{1}{(1+r_l)^2}
\Big< |{\bf \hat{q}}+{\bf \hat{q}^{\prime}} \lambda| \Big>_{\bf \hat{q}} \; .
\end{eqnarray}

Since to zeroth order in $\lambda$ Eq.~(\ref{DotSixPoint2}) determines the
flow of $v_l$ via Eq.~(\ref{eq:exactflowvl}),
 the integral on the right hand side of
Eq.~(\ref{sixpointscale}) gives simply $v_l$ to order $\lambda^0$. 
We therefore split the
angular average of the six-point vertex into two contributions,

\begin{eqnarray}
\big< \tilde{\Gamma}^{(6)}_l( {\bf{q}}_1^\prime ,
{\bf{q}}_2^\prime , {\bf{\hat{q}}}
; {\bf{\hat{q}}} , {\bf{q}}_2 , {\bf{q}}_1 ) \big>_{\bf \hat{q}} &&
 \nonumber \\
&& \hspace{-4cm}
\approx
v_l +
\big< \tilde{\Gamma}^{(6i)}_l( {\bf{q}}_1^\prime ,
{\bf{q}}_2^\prime , {\bf{\hat{q}}}
; {\bf{\hat{q}}} , {\bf{q}}_2 , {\bf{q}}_1 ) \big>_{\bf \hat{q}} \; ,
\end{eqnarray}
where $\tilde{\Gamma}^{(6i)}_l$ contains the irrelevant parts
of the six-point vertex. To calculate the angular 
average of the irrelevant
part, we can neglect
the $\eta_l$ dependence of the integral and also approximate the
retarded
dependence of the parameters on $l-t$ 
by an $l$ dependence, since the retardation
is exponentially damped.
For large $l$ we therefore can write the contributions of the
irrelevant part to
Eq.~(\ref{sixpointscale}) as follows

\begin{widetext}
\begin{eqnarray}
\big< \tilde{\Gamma}^{(6i)}_l( {\bf{q}}_1^\prime ,
{\bf{q}}_2^\prime , {\bf{\hat{q}}}
; {\bf{\hat{q}}} , {\bf{q}}_2 , {\bf{q}}_1 ) \big>_{\bf \hat{q}}
&\approx&
-v_l \int_0^1 \frac{d\lambda}{\lambda} \Big\{6 u_l (\dot{\chi}_l(\lambda)-\beta_0)\nonumber
+a_l\left(  4\dot{\chi}_l(\lambda)(1+\lambda)+4\dot{\Phi}_l(\lambda)
-8 \beta_0
+ [\dot{\varphi}_l(\lambda)-\dot{\varphi}_l(0)]
\right)
\\ && \hspace{-2.2cm} \nonumber
+b_l\left( 2  \dot{\chi}_l(\lambda)(1+\lambda)+2 \dot{\Phi}_l(\lambda)
-4 \beta_0
+  [\dot{\varphi}_l(\lambda)-\dot{\varphi}_l(0)]
\right)
+\lambda |{\bf q}_1+{\bf q}_2| \left(\beta_1 u_l +(2\beta_1+ \beta_2)a_l+\beta_0 b_l
\right)
\\ && \hspace{-2.2cm}
+\lambda \left(|{\bf q}_1-{\bf q}_1^\prime|+|{\bf q}_1-{\bf q}_2^\prime|\right)
\left( 2\beta_1 u_l+(2\beta_0+2\beta_1+\beta_2)a_l+(2\beta_1+\beta_2)b_l\right)
\Big\} \; .
\label{DotGamma6zero}
\end{eqnarray}
\end{widetext}

We now turn to the calculation of the BCS, ZS and ZS$^\prime$
diagrams in Eq.~(\ref{eq:dotGammaabcagain}). Keeping only the
relevant and marginal terms in the expansion of the four-point
vertex, see Eq.~(\ref{eq:fourpointexp}), yields all contributions to
second order in the marginal and relevant parameters arising from
these diagrams and the irrelevant part of the four-point vertex does
not contribute at this order.
\subsubsection{Contribution of the BCS, ZS and ZS$^\prime$ diagrams}
The contribution from the BCS-channel to Eq.~(\ref{eq:dotGammaabcagain}) is

\begin{eqnarray}
  \dot{\Gamma}_{l}^{(4, {\rm{BCS}} ) }
( {\bf{q}}_1^{\prime} , {\bf{q}}_2^{\prime} ; {\bf{q}}_2 , {\bf{q}}_1 )
 & \approx  &
  \nonumber
 \\
 &  & \hspace{-40mm}
 - \frac{1}{2} \left[ {u}_l
 + b_l |  {\bf{q}}_1 + {\bf{q}}_2   |
 \right]^2
    \dot{\chi}_l ( |  {\bf{q}}_1 + {\bf{q}}_2 | )
 \nonumber
 \\
  &  & \hspace{-40mm} -
 a_l \left[ u_l + b_l  |  {\bf{q}}_1 + {\bf{q}}_2 | \right]
  \nonumber \\
 & &  \hspace{-35mm}
 \times
 \frac{1}{1+r_l}
 \Big<
 \frac{
\Theta ( 1 <  |  {\bf{q}}_1 + {\bf{q}}_2 - {\bf \hat{q}}|  < e^{ l} ) }{
  |  {\bf{q}}_1 + {\bf{q}}_2 - {\bf{\hat{q}}}|^2 +r_l  }
 \nonumber
 \\
 & & \hspace{-35mm}
 \times \bigl[
 | {\bf{\hat{q}}} - {\bf{q}}_1^{\prime} |
 +  | {\bf{\hat{q}}} - {\bf{q}}_2^{\prime} |
 +
 | {\bf{\hat{q}}} - {\bf{q}}_2 |
 +  | {\bf{\hat{q}}} - {\bf{q}}_1 | \bigr]
 \Big>_{\bf \hat{q}}
 \nonumber
 \\
  &  & \hspace{-40mm} -
 \frac{a_l^2}{1+r_l}
 \Big<
 \frac{
\Theta ( 1 <  |  {\bf{q}}_1 + {\bf{q}}_2 - {\bf{\hat{q}}}|  < e^{ l} ) }{
 |  {\bf{q}}_1 + {\bf{q}}_2 - {\bf{\hat{q}}}|^2 + r_l )  }
 \nonumber
 \\
 & & \hspace{-43mm}
 \times
 \bigl(
 | {\bf{\hat{q}}}^{\prime} - {\bf{q}}_1^{\prime} |
 +  | {\bf{\hat{q}}}^{\prime} - {\bf{q}}_2^{\prime} | \bigr)
 \bigl(
 | {\bf{\hat{q}}} - {\bf{q}}_2 |
 +  | {\bf{\hat{q}}}- {\bf{q}}_1 | \bigr)
 \Big>_{\bf \hat{q}}
 ,
 \label{eq:BCSfull}
 \end{eqnarray}
and the contribution from the zero-sound channel is

\begin{eqnarray}
  \dot{\Gamma}_{l}^{(4, {\rm{ZS}} ) }
( {\bf{q}}_1^{\prime} , {\bf{q}}_2^{\prime} ; {\bf{q}}_2 , {\bf{q}}_1 )
 & \approx  &
  \nonumber
 \\
 &  & \hspace{-40mm}
 -  \left[ {u}_l
 + a_l |  {\bf{q}}_1 - {\bf{q}}_1^{\prime}   |
 \right]^2
 \dot{\chi}_l ( |  {\bf{q}}_1 - {\bf{q}}_1^{\prime} | )
 \nonumber
 \\
  &  & \hspace{-40mm} -
 ( a_l + b_l ) \left[ u_l + a_l  |  {\bf{q}}_1 - {\bf{q}}_1^{\prime} | \right]
 \nonumber
 \\
 & & \hspace{-35mm} \times
 \frac{1}{1+r_l}
 \Big<
 \frac{
\Theta ( 1 <  |  {\bf{q}}_1 - {\bf{q}}_1^{\prime} +
{\bf{\hat{q}}}|  < e^{ l} ) }{
 |  {\bf{q}}_1 - {\bf{q}}_1^{\prime} + {\bf{\hat{q}}}|^2 +r_l   }
 \nonumber
 \\
 & & \hspace{-35mm}
 \times
 \bigl[
 | {\bf{\hat{q}}} - {\bf{q}}_1^{\prime} |
 +  | {\bf{\hat{q}}}+ {\bf{q}}_2^{\prime} |
 +
 | {\bf{\hat{q}}} - {\bf{q}}_2 |
 +  | {\bf{\hat{q}}} + {\bf{q}}_1 | \bigr]
 \Big>_{\bf \hat{q}}
 \nonumber
 \\
  &  & \hspace{-40mm} -
 ( a_l^2 + b_l^2 )
 \frac{1}{1+r_l}
 \Big<
 \frac{
\Theta ( 1 <  |  {\bf{q}}_1 - {\bf{q}}_1^{\prime} + {\bf{\hat{q}}}|  < e^{ l} )
 }{
 |  {\bf{q}}_1 - {\bf{q}}_1^{\prime} + {\bf{\hat{q}}} |^2 + r_l   }
 \nonumber
 \\
 & & \hspace{-35mm}
 \times
 \bigl[
 | {\bf{\hat{q}}} - {\bf{q}}_1^{\prime} |
 | {\bf{\hat{q}}} - {\bf{q}}_2 | +
 | {\bf{\hat{q}}} + {\bf{q}}_1 |
 | {\bf{\hat{q}}} + {\bf{q}}_2^{\prime} | \bigr]
 \Big>_{\bf \hat{q}}
\nonumber
 \\
  &  & \hspace{-40mm} -
 2 a_l b_l
 \frac{1}{1+r_l}
 \Big<
 \frac{
\Theta ( 1 <  |  {\bf{q}}_1 - {\bf{q}}_1^{\prime} + {\bf{\hat{q}}}|  < e^{ l} )
 }{
 |  {\bf{q}}_1 - {\bf{q}}_1^{\prime} + {\bf{\hat{q}}} |^2 + r_l   }
 \nonumber
 \\
 & & \hspace{-35mm}
 \times
 \bigl[
 | {\bf{\hat{q}}} - {\bf{q}}_1^{\prime} |
 | {\bf{\hat{q}}} + {\bf{q}}_2^{\prime} | +
 | {\bf{\hat{q}}} + {\bf{q}}_1 |
 | {\bf{\hat{q}}} - {\bf{q}}_2 | \bigr]
 \Big>_{\bf \hat{q}}
 \; .
 \label{eq:ZSfull}
 \end{eqnarray}

The contribution from the other zero-sound  channel (ZS$^{\prime}$
in Fig.~\ref{fig:FourPoint})
is obtained by replacing ${\bf{q}}_1^{\prime} \leftrightarrow {\bf{q}}_2^{\prime}$
on the right-hand side of Eq.~(\ref{eq:ZSfull}).
Expanding to linear order in the momenta, we arrive at
 \begin{eqnarray}
  \dot{\Gamma}_{{\rm lin},l}^{(4, {\rm{BCS}} ) }
( {\bf{q}}_1^{\prime} , {\bf{q}}_2^{\prime} ; {\bf{q}}_2 , {\bf{q}}_1 )
 & \approx  & - \beta_0 \left[ \frac{u_l^2}{2}  + 2 a_l^2  + 2 u_l a_l \right]
 \nonumber
 \\
  &  & \hspace{-40mm} -  | {\bf{q}}_1 + {\bf{q}}_2 |
 \Bigl[ \frac{ \beta_1}{2} u_l^2 + 2 (  \beta_1 +  \beta_2 ) a_l^2
 + ( 2 \beta_1 + \beta_2 ) u_l a_l
 \nonumber
 \\
 & & \hspace{-21mm}
  + \beta_0 u_l b_l + 2 \beta_0  a_l b_l \Bigr]
 \; .
 \label{eq:BCSflow}
 \end{eqnarray}

Similarly, we obtain for the contribution from  the
zero-sound-channel,
 \begin{eqnarray}
  \dot{\Gamma}_{{\rm lin},l}^{(4, {\rm{ZS}} ) }
( {\bf{q}}_1^{\prime} , {\bf{q}}_2^{\prime} ; {\bf{q}}_2 , {\bf{q}}_1 )
 & \approx  & - \beta_0 \left[ u_l +  a_l + b_l \right]^2
 \nonumber
 \\
  &  & \hspace{-40mm} -  | {\bf{q}}_1 - {\bf{q}}_1^{\prime} |
 \Bigl[  \beta_1  u_l^2 + ( 2 \beta_0 + \beta_1 +  \beta_2 ) a_l^2
 + (  \beta_1 + \beta_2 ) b_l^2
 \nonumber
 \\
 & & \hspace{-21mm}
  + ( 2 \beta_0 + 2 \beta_1 + \beta_2 ) u_l a_l  + ( 2 \beta_1 + \beta_2 ) u_l b_l
 \nonumber
 \\
 & & \hspace{-21mm}
 +2  (  \beta_0 + \beta_1 +  \beta_2 )  a_l b_l \Bigr]
 \; .
 \label{eq:ZSflow}
 \end{eqnarray}

The contribution from ZS$^\prime$ is again obtained by replacing
${\bf q}_1^\prime \leftrightarrow {\bf q}_2^\prime$.

\subsection{Truncated flow equation for the two-point vertex}

To calculate the flow of $r_l$, we need to calculate
$\dot{\Gamma}^{(2)}_l(0)$, see Eq.~(\ref{eq:exactflowrl}). We write
it as

\begin{widetext}
\begin{eqnarray}
\label{gamma2von0}
(1+r_l)\dot{\Gamma}^{(2)}_l(0)&=&
\left<\tilde\Gamma_l^{(4)}
(0,{\bf \hat{q}},{\bf \hat{q}},0)
\right>_{ {\bf \hat{q}}}
\approx e^{l-2 \int_0^l d\tau \eta_\tau}u_0
+
\int_0^l dt e^{t-2 \int_{l-t}^l d\tau \eta_\tau}
\left<
\dot{\Gamma}_{l-t}^{(4)}(0,e^{-t}
{\bf \hat{q}},e^{-t} {\bf \hat{q}},0)
\right>_{ {\bf \hat{q}}}
\end{eqnarray}
with the second order expression of $\dot{\Gamma}_l^{(4)}$ given by
Eq.~(\ref{eq:dotGammaabcagain}).
Thus, we have
\begin{eqnarray}
\label{gamma4punktav}
\dot{\Gamma}_l^{(4)}(0,\lambda {\bf \hat{q}},\lambda {\bf \hat{q}},0)
&\approx&
-\frac{1}{2}\dot{\chi}_l(\lambda)
\big[(u_l+\lambda b_l+a_l)^2+(u_l+b_l+a_l \lambda)^2+(u_l +a_l[1+\lambda])^2
\big]
\nonumber \\ &&
-\dot{\phi}_l(\lambda)
\big[a_l(u_l+b_l \lambda + a_l)
+a_l(u_l+b_l+a_l \lambda) + b_l(u_l+a_l[1+\lambda]) \big]
\nonumber \\
&&
-\frac{1}{2} \dot{\psi}_l(\lambda)
\big[2 a_l^2 +b_l^2]
- (u_l+b_l+a_l)\left[ \beta_0 u_l+(b_l+a_l)
\dot{\varphi}_l(\lambda)
\right]
\nonumber \\ &&
+\frac{1}{1+r_l}\big<
\tilde{\Gamma}^{(6)}_l(0,\lambda {\bf \hat{q}},{\bf \hat{q}}^\prime;
{\bf \hat{q}}^\prime,\lambda {\bf \hat{q}},0) \big>_{{\bf \hat{q}}^\prime}
\; ,
\end{eqnarray}
\end{widetext}
with
\begin{eqnarray}
\dot{\psi}_l(\lambda)=\frac{2}{1+r_l} \Big< \tilde{G}_l({\bf \hat{q}}+{\bf \hat{q}}^\prime \lambda)
|{\bf \hat{q}}+{\bf \hat{q}}^\prime \lambda|^2 \Big>_{\bf \hat{q}} \; .
\end{eqnarray}
Keeping only terms up to linear order in $\lambda$, which define the
flow of
$u_l$, $a_l$ and $b_l$,
one finds
\begin{eqnarray}
\dot{\Gamma}^{(2)}_l(0) &\approx& \frac{u_l+a_l+b_l}{1+r_l},
\label{gamma2punktapprox1}
\end{eqnarray}
which is just the contribution of the relevant and marginal parts
of $\big<\tilde\Gamma_l^{(4)}
(0,{\bf \hat{q}},{\bf \hat{q}},0)
\big>_{ {\bf \hat{q}}}$. Higher orders in $\lambda$ correspond
to the contribution of the irrelevant part of the four-point vertex.
With
\begin{eqnarray}
\big<\tilde\Gamma_l^{(4)}
(0,{\bf \hat{q}},{\bf \hat{q}},0)
\big>_{ {\bf \hat{q}}}&=&
u_l+a_l+b_l
\nonumber \\ &&
+\big<\tilde\Gamma_l^{(4 \rm i)}
(0,{\bf \hat{q}},{\bf \hat{q}},0)
\big>_{ {\bf \hat{q}}}
\end{eqnarray}
we can write the contributions of the irrelevant parts (we here
ignore the contributions of the irrelevant part of the six-point vertex
beyond those which are implicitly contained in the renormalization
of $u_l$) as
\begin{eqnarray}
\big<\tilde{\Gamma}^{(4 \rm i)}_l(0,{\bf \hat{q}},{\bf \hat{q}},0)
\big>_{ {\bf \hat{q}}} &\approx&
\int \frac{d \lambda}{\lambda^2}
\Big<
\dot{\Gamma}^{(4,{\rm BCS})}_l
(0,\lambda {\bf \hat{q}};\lambda {\bf \hat{q}},0)
\nonumber \\ &&
\hspace{-2.2cm}
+\dot{\Gamma}^{(4,{\rm ZS})}_l
(0,\lambda {\bf \hat{q}};\lambda {\bf \hat{q}},0)
+\dot{\Gamma}^{(4,{\rm ZS}^\prime)}_l
(0,\lambda {\bf \hat{q}};\lambda {\bf \hat{q}},0)
\nonumber \\ &&
\hspace{-2.2cm}
-\dot{\Gamma}^{(4,{\rm BCS})}_{{\rm lin},l}
(0,\lambda {\bf \hat{q}};\lambda {\bf \hat{q}},0)
-\dot{\Gamma}^{(4,{\rm ZS})}_{{\rm lin},l}
(0,\lambda {\bf \hat{q}};\lambda {\bf \hat{q}},0)
\nonumber \\ &&
\hspace{-2.2cm}
-\dot{\Gamma}^{(4,{\rm ZS}^\prime)}_{{\rm lin},l}
(0,\lambda {\bf \hat{q}};\lambda {\bf \hat{q}},0)
\Big>_{\bf \hat{q}} \; ,
\label{gamma2punktapprox2}
\end{eqnarray}
which is identical to the second and higher order in $\lambda$ contributions
contained in Eq.~(\ref{gamma4punktav}).

\subsection{Flow equations of marginal and relevant parameters}

It is now straightforward to write down the flow equations for
the marginal and relevant coupling parameters. The flow of $v_l$
is determined by Eq.~(\ref{eq:exactflowvl}).
With Eq.~(\ref{eq:sixpointapprox2}) one finds
 \begin{eqnarray}
\partial_l v_l & = & ( 6 - 2 D - 3 \eta_l ) v_l  -  3 \beta_0 v_l( 4 u_l
+5 a_l +3 b_l)
\nonumber \\
&& \hspace{-.3cm}
- \frac{4 v_l^2}{1+r_l} \int_0^1 d \lambda \ \lambda^{4D-9}
\big[8 \dot{\chi}_{l}(0)+9\dot{\chi}_{l}(\lambda)
\big] \; ,
 \label{eq:vfloweta}
\end{eqnarray}
where we approximated the retarded dependence on
$v_{l-t}$ and $\dot{\chi}_{l-t}$ in Eq.~(\ref{eq:sixpointapprox2})
by $v_{l}$ and $\dot{\chi}_{l}$ (since the retardation is exponentially
damped) and further took for the lower bound of the integral the
limit $l\to \infty$.
Eq.~(\ref{eq:vfloweta}) obviously has a $v_*=0$
fixed point solution, which is stable if
$3\beta_0(4u_*+5a_*+3 b_*)>6-2D-3 \eta$.

Collecting all contributions to Eq.~(\ref{eq:dotGammaabcagain}) from
Eqs.~(\ref{DotGamma6zero}, \ref{eq:BCSflow}, \ref{eq:ZSflow}), and
decomposing $\dot{\Gamma}^{(4)}_l
({\bf q}_1^\prime,{\bf q}_2^\prime;{\bf q}_2,{\bf q}_1)$ into a
momentum independent contribution and linear contributions according
to Eq.~(\ref{eq:dotGammaabcdef}), we find, using
Eqs.~(\ref{eq:exactflowul}, \ref{eq:flowa}, \ref{eq:flowb}), the flow
equations for the parameters characterizing the four-point vertex,
 \begin{eqnarray}
 \partial_l u_l
 & = & ( 4 - D - 2 \eta_l ) u_l - \frac{5}{2} \beta_0 u_l^2
 -   \beta_0 u_l (6 a_l  +  4 b_l)
 \nonumber
 \\
& & \hspace{-.5cm} - 2 \beta_0 (2 a_l^2 + b_l^2 +2 a_l b_l) +
\frac{v_l}{ 1 + r_l} -\frac{v_l}{1+r_l} \int_0^1
\frac{d\lambda}{\lambda}
 \nonumber \\ && \times \Big\{ 6 u_l
(\dot{\chi}_l(\lambda)-\beta_0)\nonumber +a_l\Big(
4\dot{\chi}_l(\lambda)(1+\lambda)+4\dot{\Phi}_l(\lambda) \nonumber
\\ && \ \ \ \ -8 \beta_0 +
[\dot{\varphi}_l(\lambda)-\dot{\varphi}_l(0)] \Big) +b_l\Big( 2
\dot{\chi}_l(\lambda)(1+\lambda) \nonumber \\ && \ \ \ \ +2
\dot{\Phi}_l(\lambda) -4 \beta_0 +
[\dot{\varphi}_l(\lambda)-\dot{\varphi}_l(0)] \Big) \Big\}
 \; ,
 \label{eq:uflow2eta}
\end{eqnarray}

\begin{eqnarray}
 \partial_l a_l & = & ( 3 - D - 2 \eta_l ) a_l -  \beta_1 u_l^2
 -    (2 \beta_0 + 2 \beta_1 + \beta_2  ) u_l a_l
\nonumber \\ &&
- ( 2 \beta_1 +  \beta_2) u_l b_l
- ( 2 \beta_0 + \beta_1 + \beta_2) a_l^2
\nonumber \\ &&
- ( \beta_1 + \beta_2)  b_l^2
- 2 ( \beta_0 + \beta_1 + \beta_2 ) a_l b_l
\nonumber \\ &&
-\frac{v_l}{1+r_l}\big\{ 2 \beta_1 u_l+
\left(2 \beta_0+2\beta_1+\beta_2 \right) a_l
\nonumber \\ &&
\hspace{1.35cm} + \left(2 \beta_1 +\beta_2 \right) b_l
\big\}
 \; ,
 \label{eq:aflow2eta}
\\
 \partial_l b_l & = & ( 3 - D - 2 \eta ) b_l -  \frac{\beta_1}{2} u_l^2
 -    ( 2 \beta_1 + \beta_2  ) u_l a_l
\nonumber \\ &&
 -  \beta_0 u_l b_l
  -2 (  \beta_1 + \beta_2) a_l^2
 - 2  \beta_0   a_l b_l
\nonumber \\ &&
-\frac{v_l}{1+r_l}\big\{ \beta_1 u_l+
\left(2\beta_1+\beta_2 \right) a_l +\beta_0 b_l \big\}
 \; .
 \label{eq:bflow2eta}
\end{eqnarray}
The flow equation for $r_l$ is obtained from Eq.~(\ref{eq:exactflowrl})
and Eqs.~(\ref{gamma2von0}-\ref{gamma2punktapprox2}),
\begin{eqnarray}
 \partial_l r_l &=& ( 2 - \eta_l ) r_l +
 \frac{ u_l + a_l + b_l}{ 1  + r_l }
\nonumber \\ &&
 + \frac{1}{1+r_l}
 \big<
 \tilde{\Gamma}^{(4 \rm i)}_l(0,{\bf \hat{q}};{\bf \hat{q}},0)
 \big>_{\bf \hat{q}}
 \; .
 \label{eq:rflowmargeta}
 \end{eqnarray}
Finally, we need to determine the flowing anomalous dimension $\eta_l$
before we can analyse the fixed point of the RG equations.
For now, we keep
only the marginal and relevant part of the four-point vertex in
Eq.~(\ref{eq:anomalexplicit}) and arrive at
\begin{eqnarray}
 \eta_l & \approx &  \beta_3 (a_l + b_l)
 \; ,
 \label{eq:etares}
 \end{eqnarray}
where
 \begin{equation}
 \beta_3 =
\frac{1}{1+r_l}
  \frac{ \partial}{\partial q^2 }
\big<
        |  {\bf{\hat{q}^\prime}} + {\bf{q}}  |
\big>_{\bf \hat{q}^\prime} \Big|_{ q^2 =0}
 = \frac{D-1}{2D (1+r_l)}
 \label{eq:beta3def}
 \; .
 \end{equation}

\subsubsection{Fixed point values in $D=3$}
\label{FPsection}
In $D=3$, we have
\begin{eqnarray}
\dot{\Phi}_l(\lambda)&=& \frac{\lambda+
\sqrt{-r_l}\ \mbox{arctanh}\left[\frac{\lambda \sqrt{-r_l}}{1+r_l+\lambda}
\right]}
{\lambda (1+r_l)} \, ,
\\
\dot{\psi}_l(\lambda)&=&\frac{\lambda (2+\lambda)+r_l \ln
\left[\frac{1+r_l}{(1+\lambda)^2+r_l}\right]
}
{2\lambda (1+r_l)} \, ,
\\
\dot{\varphi}_l(\lambda) &=& \frac{3+\lambda^2}{3 (1+r_l)^2} \, ,
\end{eqnarray}
and 
\begin{subequations}
\begin{eqnarray}
\beta_0 &=& \frac{1}{(1+r_l)^2} \; , \\ 
\beta_1 &=& \frac{r_l-1}{2 (1+r_l)^3} \; , \\
\beta_2 &=& \frac{1}{2 (1+r_l)^2} \; , \\
\beta_3 &=&  \frac{1}{3(1+r_l)} \; .
\end{eqnarray}
\end{subequations}
Setting the left-hand sides of the
five  flow equations
 (\ref{eq:vfloweta},\ref{eq:uflow2eta},\ref{eq:aflow2eta},\ref{eq:bflow2eta},\ref{eq:rflowmargeta})
for the two relevant coupling parameters
 $r_l$ and $u_l$ and the three marginal coupling parameters
$a_l$, $b_l$ and $v_l$ equal to zero and employing
Eq.~(\ref{eq:etares}) for the 
anomalous dimension, we obtain numerically the fixed
point values
\begin{equation}
    \begin{array}{lll}
    r_{\ast}  \approx  -0.0996
    \; , &
 u_{\ast}  \approx 0.122
 \; ,  &
 a_{\ast}  \approx  0.0371
 \; , \\
 b_{\ast}  \approx  0.0339
 \; ,
& v_{\ast}  =  0
 \; ,
& \eta  \approx  0.0263\ .
\end{array}
 \label{eq:fixpoints0}
 \end{equation}
The value of $\eta$ is now much closer to the correct value
$\eta\approx 0.038$ and the inclusion of the marginal terms
certainly improves upon the analysis including only $u_l$.
However, the fact that $v_*=0$ is an artifact of the
approximation which ignores all terms of third or higher order
in the relevant and marginal parameters. A simple improvement
can be obtained by including from the third order terms in
the renormalization of $v_l$ the marginal contributions. In that case, only the
flow of $v_l$ is modified and Eq.~(\ref{eq:vfloweta})
becomes
\begin{eqnarray}
  \partial_l v_l & = & ( 6 - 2 D - 3 \eta_l ) v_l  -  3 \beta_0 v_l( 4 u_l
  +5 a_l +3 b_l)
  \nonumber \\
  && - \frac{4 v_l^2}{1+r_l} \int_0^1 d \lambda \ \lambda^{4D-9}
  \big[8 \dot{\chi}_{l}(0)+9\dot{\chi}_{l}(\lambda)
  \big] \;
  \nonumber \\
  & + &   \frac{ ( u_l + a_l + b_l )     }{ (1 + r_l )^3}
  \Bigl[
  12 ( u_l + a_l + b_l )^2
  \nonumber \\
  & &
  \hspace{25mm} + 9 ( u_l + 2 a_l )^2
  \Bigr]
  \; .
  \label{eq:vfloweta:3rdorder}
\end{eqnarray}
The resulting fixed point values with the 3rd order terms are:
\begin{equation}
    \begin{array}{lll}
     r_{\ast}  \approx  -0.134 \; , &
     u_{\ast}  \approx  0.127 \; ,  &
     a_{\ast}  \approx  0.0693 \; , \\
     b_{\ast}  \approx  0.0639 \; , &
     v_{\ast}  \approx  0.0917 \; , &
     \eta  \approx  0.0513\; .
    \end{array}
  \label{eq:fixpoints1}
\end{equation}
The value for $\eta$ is now slightly too large and a quick
convergence is not obtained.
However, one may expect from this
result that a consistent treatment to third order, a
very complex calculation, would indeed improve upon the second
order result. For completeness, let us also mention the results
of another possible approximation, where one keeps only the marginal
and relevant terms of the vertices entering the $\dot{\Gamma}^{(n)}_l$
expressions. Within such an approximation,
the flow equations for $a_l$ and $b_l$
remain identical to Eqs.~(\ref{eq:aflow2eta},\ref{eq:bflow2eta}) whereas
the flow equations for $u_l$, $v_l$ and $r_l$ would simplify to
\begin{eqnarray}
  \partial_l v_l & = & ( 6 - 2 D - 3 \eta_l ) v_l  -  3 \beta_0 v_l( 4 u_l
  +5 a_l +3 b_l)
  \nonumber \\
  & + &   \frac{ ( u_l + a_l + b_l )     }{ (1 + r_l )^3}
  \Bigl[
  12 ( u_l + a_l + b_l )^2
  \nonumber \\
  & &
  \hspace{25mm} + 9 ( u_l + 2 a_l )^2
  \Bigr]
  \; ,
  \label{simpleflowv}
  \\
 \partial_l u_l
 & = & ( 4 - D - 2 \eta_l ) u_l - \frac{5}{2} \beta_0 u_l^2
 -   \beta_0 u_l (6 a_l  +  4 b_l) 
 \nonumber
 \\
 & &
 \hspace{-.5cm}
 - 2 \beta_0 (2 a_l^2 + b_l^2 +2 a_l b_l)
 + \frac{v_l}{ 1 + r_l} \; ,
 \\
 \partial_l r_l &=& ( 2 - \eta_l ) r_l +
 \frac{ u_l + a_l + b_l}{ 1  + r_l } \; .
\end{eqnarray}
The corresponding fixed point values are
 \begin{equation}
    \begin{array}{lll}
     r_{\ast}  \approx  -0.227 \; , &
     u_{\ast}  \approx  0.178 \; , &
     a_{\ast}  \approx  0.0838 \; , \\
     b_{\ast}  \approx  0.0767 \; , &
     v_{\ast}  \approx  0.255 \; , &
     \eta  \approx  0.0692 \ .
    \end{array}
 \label{eq:fixpoints2}
 \end{equation}
The value for $\eta$ is worse in this approximation compared to
those which also include irrelevant parts of the vertices. Problems
with a similar implementation of the sharp cutoff formulation were
previously reported in Ref.~\cite{Morris96}. Ignoring the
third order terms in the flow of $v_l$, Eq.~(\ref{simpleflowv}),
leads to fixed point values almost identical to those listed in
Eqs.~(\ref{eq:fixpoints0}).

\subsubsection{Including irrelevant terms in the equation for
$\eta$}
\label{subsect:etairr}
Eq.~(\ref{eq:etares}) for the flowing anomalous dimension includes
only the marginal parts of the four-point vertex. To include also
irrelevant contributions
we need to evaluate
the irrelevant parts of $\dot{\Gamma}^{(4)}_l$, i.~e.~we must calculate
both the irrelevant contribution from the six-point vertex and
the BCS, ZS and ZS$^\prime$ contributions to
\begin{eqnarray}
\label{eq:etaPSA} \eta &=& \frac{1}{1+r_*}\int_0^1 d\lambda
\lambda^{-2+2\eta}
\nonumber \\
&&  \hspace{.4cm} \times \frac{\partial}{\partial q^2} \left. \Big<
\dot{\Gamma}^{(4)}_\infty(\lambda{\bf q},\lambda{\bf \hat{q}^\prime},\lambda
{\bf \hat{q}^\prime},\lambda{\bf q})\Big>_{\bf \hat{q}^\prime}\right|_{q^2=0}
\end{eqnarray}
using
Eqs.~(\ref{Gamma6punkt},\ref{sixpointscale},\ref{DotGamma6zero},\ref{eq:BCSfull},\ref{eq:ZSfull}).
This is a somewhat lengthy calculation, and we refer to Appendix
\ref{App:eta} for a table of the required integrals. The resulting
values of $\eta$ are generally very small, in the approximation with
all terms up to second order in the relevant and marginal parameters
we obtain $\eta\approx 0.0127$. Including the third order terms in
the flow of $v_l$ we get the even smaller value
$\eta\approx0.00797$. It is not completely clear why the results for
$\eta$ become worse on including irrelevant terms in the calculation
for $\eta_l$. A possible problem is that in the calculation of
$\eta$, Eq.~(\ref{eq:etaPSA}), irrelevant terms contribute which
however do not also enter the renormalization of $u_l$, $a_l$ and
$b_l$. Terms similar to the irrelevant ones contained in
Eq.~(\ref{eq:etaPSA}) would only enter the renormalization of other
parameters at third order in the marginal and relevant parameters.
This might suggest that one should use the same level of approximation
for the vertex $\tilde{\Gamma}^{(4)}_l$ in calculating
$\eta$ and $\dot{\Gamma}^{(4)}_l$ for the
flow equations for $u_l$, $a_l$ and $b_l$. In that
case, the first order expression for $\eta$, Eq.~(\ref{eq:etares}),
should be used together with the second order expressions for
the flows of $u_l$, $a_l$ and $b_l$. As shown in the previous
subsection, this indeed leads to better approximations of $\eta$.

\section{Calculation of $\sigma(x)$ including only marginal parameters}
\label{sec:sigmamarg}

We demonstrate here the importance of the irrelevant
parts of the four-point vertex for a correct description of the
large wave vector regime and show that a treatment that ignores irrelevant
terms leads to an incorrect description of
$\sigma(x)$ in the regime $x\gg 1$ and the
wrong result $\Delta T_c \propto u_0 \ln u_0$.
Ignoring in Eq.~(\ref{eq:fourpointexp}) the irrelevant contributions,
one
obtains from Eq.~(\ref{eq:gammadot2def})
the following expression for the subtracted function
$\dot{\Gamma}_l^{(2m)}({\bf q})=\dot{\Gamma}_l^{(2)}({\bf q})-
\dot{\Gamma}_l^{(2)}(0)$ (the superscript $m$ indicates that
only marginal terms remain)
\begin{eqnarray}
  \dot{\Gamma}_l^{(2m)}({\bf q})&\approx & \frac{a_l+b_l}{1+r_l}
\left(\left< |{\bf \hat{q}^\prime}+{\bf q}| \right>_{\bf \hat{q}^\prime}-1\right)
\nonumber \\
&=&\frac{a_l+b_l}{1+r_l} \times \left\{
\begin{array}{cl}
\frac{q^2}{3} & \mbox{for} \, q<1 \, , \\
\frac{3q^2-3 q +1}{3q}  & \mbox{for} \, q>1 \, .
\end{array}
\right.
\end{eqnarray}
Via Eq.~(\ref{eq:sigmaexact}) we obtain the scaling function
$\sigma(x)$ (we here restrict the discussion to $D=3$),
\begin{eqnarray}
\label{sigmamarg}
&& \hspace{-1.cm}\sigma(x) \approx  x^2 \int_0^{l_c-\ln x} dl \
e^{-2(l-l_c)+\int_0^l d\tau
\eta_\tau} \frac{a_l+b_l}{3(1+r_l)}
\nonumber \\ && \hspace{0.1cm}
+ \int_{l_c-\ln x}^\infty  dl \
e^{\int_0^l d\tau \,
\eta_\tau} \frac{a_l+b_l}{3(1+r_l)}  \nonumber \\
&& \hspace{.5cm}   \times \Big[
e^{-(l-l_c)}x - e^{-2(l-l_c)} +\frac{1}{3x} e^{-3(l-l_c)} \Big] .
\end{eqnarray}
This equation is valid for $\ln x<l_c$, for $\ln x\ge l_c$ the
integral bound $l_c-\ln x$ is replaced by zero, i.~e.~the first
integral vanishes. Since in $D=3$ we have
$(a_l+b_l)/(3[1+r_l])\approx \eta_l$ (see Eq.~(\ref{eq:etares})),
$\sigma(x)$ is determined by the flow of $\eta_l$ alone in this
approximation. Let us examine first the case $x \gg 1$. The first
integral then extends only over small $l\ll l_c$. In that case, all
marginal parameters are small and we may ignore the contribution of
the integral of $\eta_\tau$ in the exponent and approximate the flow
of the marginal parameters  as $a_l$, $b_l\propto e^{2(l-l_c)}
u_*^2$ since $\partial_l a_l\approx 2 \partial_l b_l \approx u_l^2
\propto e^{2(l-l_c)} u_*^2$ for $l\ll l_c$. The contribution of
the first integral in Eq.~(\ref{sigmamarg}) therefore vanishes like
$x^2(l_c-\ln x)u_*^2$ as $x$ approaches the UV cutoff, $x \to 
e^{l_c}$ and makes no contribution for $x>e^{l_c}$. To investigate the
contribution of the second integral, we split it into two parts,
\begin{equation}
\int_{l_c-\ln x}^\infty =\int_{l_c-\ln x}^{l_c} + \int_{l_c}^\infty \, ,
\end{equation}
where, to estimate the contribution of the regime $l<l_c$, we use the
same approximation as for the first integral in Eq.~(\ref{sigmamarg}). This
yields a contribution $\propto (x-1-\ln x-(1-x)/3x)$. To estimate the
contribution of the integral for $l>l_c$ we may replace the parameters
$a_l$, $b_l$, $r_l$ and $\eta_l$ by their fixed point values
and obtain a contribution $\propto \eta x/(1-\eta)-\eta/(2-\eta)+
\eta/(3x(3-\eta))$. We have employed Eq.~(\ref{eq:etares}) to arrive
at this result.
 Hence, the dominant behavior of $\sigma(x)$
for large $x$ is linear in $x$. This large $x$ behavior prohibits the
removal of the UV cutoff in the integral determining the temperature
shift, Eq.~(\ref{eq:tcshift}), and one thus would incorrectly predict from
Eq.~(\ref{eq:tcshift}) a term for the $T_c$ shift behaving
like $u_0 \ln u_0$ in $D=3$. Thus, inclusion of the irrelevant terms
is essential for determining correctly the shift of the critical temperature.
The absence of irrelevant terms in the Wilson RG analysis
carried out by Bijlsma and Stoof \cite{Bijlsma96} seems to be 
responsible for the  $u_0 \ln u_0$ behavior they find for the
$T_c$-shift.

On the other hand, an approach which explicitly includes marginal
terms but ignores irrelevant ones is well suited to describe
the scaling regime $x\to 0$. In this case, the dominant
contributions to Eq.~(\ref{sigmamarg}) come from $l>l_c$ where
we replace the coupling constants by their fixed point values
and obtain $\sigma(x)\approx A_3 x^{2-\eta}$ with
\begin{eqnarray}
\label{A3marg}
 A_3=1+\frac{3\eta}{1-\eta}
-\frac{3 \eta}{2-\eta}+\frac{\eta}{3-\eta}\approx 1+\frac{11}{6} \eta \, .
\end{eqnarray}
This expression is in relatively good agreement with our result
from Sec.~\ref{secscaling},
Eq.~(\ref{firstA3}), where we obtained
 $A_3\approx 1.17$ for $\eta\approx 0.104$,
whereas Eq.~(\ref{A3marg}) would predict  $A_3\approx 1.22$ for
the same $\eta$. Furthermore, for the generally accepted value
$\eta\approx 0.038$ Eq.~(\ref{A3marg}) predicts $A_3\approx 1.07$,
which compares relatively well with a recent Monte-Carlo
 result \cite{Prokofev04}
$A_3\approx 1.04$.

\section{Conclusion}
\label{sec:conclusions}

In this work we have shown how the functional RG can be employed to calculate
the complete scaling function of the zero frequency self-energy
of weakly interacting bosons in $3\le D<4$. The scaling
function describes the cross-over from the small wave vector
regime with anomalous scaling to the large wave vector regime
where logarithmic divergences appear in $D=3$. A simple truncation
of the flow equations at the four-point vertex which ignores the
six-point vertex and does not treat the marginal parts
of the four-point vertex consistently leads nonetheless to
a very accurate description of both the cross-over regime and
the large wave vector regime while giving satisfying
results for the anomalous scaling regime even in $D=3$. We have used this
scaling function to calculate the interaction induced 
shift of the critical temperature
and obtained $\Delta T_c/T_c = 1.23\, a n^{1/3}$,
a result which compares very well with those obtained
within the variational perturbation theory \cite{Kleinert03,Kastening03}
and Monte-Carlo simulations  \cite{MC01a,MC01b}.
The technique and truncation scheme employed here might also prove
useful to obtain energy and/or momentum-dependent scaling functions
in other critical systems. A similar truncation has already been
applied to calculate the self-energy in the vicinity of the
Luttinger liquid fixed point of fermions in one dimension \cite{Busche02}.

We have further investigated the fixed point structure using several
approximation schemes which include the marginal terms associated
with the four-point and six-point vertices in $D=3$. These schemes
generally lead to an improved anomalous dimension, however, we did
not get a quick convergence of $\eta$ to the accepted value.
Nonetheless, the best value for $\eta$ which we obtain, $\eta\approx
0.0513$ is rather close to the one obtained from the first order
average action approximation, $\eta\approx 0.049$ \cite{Wetterich01}
and it is feasible that an improved treatment, e.~g.~a consistent
treatment to third order in the marginal and relevant parameters,
would indeed produce quite accurate results. Unfortunately, such a
calculation is rather lengthy and seems not to be
an efficient way of calculating accurate values of critical
exponents.

Finally, we have analysed the two-point scaling function within an
approach which includes marginal and ignores irrelevant terms
of the four-point function. It was shown, that if one ignores the
irrelevant terms contained in the higher-order momentum 
dependence of the four-point vertex, one obtains an 
incorrect UV behavior of
the scaling function and hence the wrong functional dependence
$\Delta T_c/T_c \propto a n^{1/3} \ln (a n^{1/3})$
of the critical temperature shift in $D=3$. However, the approach worked
well in the critical regime where we used it to
calculate the prefactor of the anomalous scaling term, which
was shown to be in good agreement with numerical results \cite{Prokofev04}.
While we have not attempted to calculate the scaling function
including both marginal and irrelevant terms, it seems certainly
feasible to do so within an approximation which ignores
the six-point vertex (since $v_*=0$ in a calculation to second order
in $u_l$, $a_l$, $b_l$ and $v_l$, this would  be 
consistent). The fixed point value of $\eta$ and hence the
description of the small wave vector regime
 improves within such an approach. However, we do not expect
that the inclusion of marginal terms would have much effect
on  the crossover regime $x\approx 1$, since this regime
seems to be well described already in the simpler truncation
used in Sec.~\ref{SECself},
as is evident from the rather accurate value for the
$T_c$-shift.


\begin{appendix}
\section{Flow equations for the six- and eight-point vertex}
\label{AppMarg}
 Below, we discuss the flow equations of the  six-
and eight-point vertex for the unrescaled vertex functions. The
inhomogeneous part of the flow equations for the classical rescaled
flow equations follow via the replacement rule $\dot{G}_{\Lambda}
(K) \rightarrow - \dot{G}_l ( {\bf{q}} )$, $G_{\Lambda} (K)
\rightarrow - \tilde{G}_l ( {\bf{q}} )$, $\Gamma^{(2n)}_{\Lambda}
 \rightarrow \tilde{\Gamma}^{(2n)}_l$, $\int_K \rightarrow
\int_{\bf{q}}$, and multiplying
the resulting expression by an overall minus sign.
\subsection{Six-point vertex}
For the calculation of the critical exponents in $D=3$ we  need 
the six-point vertex, see Sec.~\ref{sec:self3}.
We first define the symmetrization operators
${\cal{S}}_{1,2,3}$ and ${\cal{S}}_{ 1,(2,3) }$ as follows,
\begin{subequations}
 \begin{eqnarray}
 {\cal{S}}_{ 1,2,3} f ( 1,2,3 ) & = &
 \frac{1}{6} \left[ f ( 1,2,3) + f ( 2,3,1) + f (3,1,2)
 \right.
 \nonumber
 \\
 & & \hspace{-11mm} \left.
+ f (3,2,1) +
 f ( 2,1,3 ) + f ( 1,3,2 ) \right]
\; ,
 \label{A123a}
 \end{eqnarray}

 \begin{eqnarray}
  {\cal{S}}_{ 1,(2,3)} f ( 1,2,3 ) & =&
 {\cal{S}}_{ (2,3), 1} f ( 1,2,3 )
   \nonumber \\ && \hspace{-15mm} =
 \frac{1}{3} \left[ f ( 1,2,3) + f ( 2,1,3) + f (3,2,1) \right] \; .
 \label{A123b}
 \end{eqnarray}
\end{subequations}
Given a function $f (1,2,3)$ that is already
symmetric with respect to the pair $(2,3)$, the function
${\cal{S}}_{ 1,(2,3)} f (1,2,3)$
is a totally symmetric function.
The flow equation of the six-point vertex is given by \cite{Busche01}
(the equation is shown graphically in Fig.~\ref{fig:SixPoint})
\begin{widetext}
 {\small
 \begin{eqnarray}
 \partial_\Lambda \Gamma^{(6)}_{\Lambda }
 ( K_1^{\prime} , K_2^{\prime} , K_3^{\prime} ;
 K_3 , K_2  , K_1 )
 & = &    \int_K
 \dot{G}_{\Lambda} ( K )
 \Gamma^{(8)}_{ \Lambda }
 ( K_1^{\prime} , K_2^{\prime} , K_3^{\prime} ,
 K ; K , K_3 ,  K_2 , K_1 )
 \nonumber
 \\
 & & \hspace{-53mm } + 3  \int_K  \Bigl\{  {\cal{S}}_{3, (2,1)}
 \Bigl[
 \dot{G}_{\Lambda} ( K )   G_{\Lambda} ( K^{\prime} )
  \Gamma^{(6)}_{ \Lambda }
 ( K_1^{\prime} , K_2^{\prime}, K_3^{\prime} ;
 K_3 , K^{\prime} , K )
 \Gamma^{(4)}_{ \Lambda }
 ( K , K^{\prime} ; K_2 , K_1 )
 \Bigr]_{ K^{\prime} = K_1 + K_2 - K}
 \nonumber
 \\
 &  &
\hspace{-47mm } +  {\cal{S}}_{ (1^{\prime},2^{\prime}),3^{\prime}}
 \Bigl[
 \dot{G}_{\Lambda} ( K )   G_{\Lambda } ( K^{\prime} )
 \Gamma^{(4)}_{ \Lambda }
 ( K_1^{\prime} , K_2^{\prime} ; K^{\prime} , K )
 \Gamma^{(6)}_{ \Lambda }
 ( K , K^{\prime} , K_3^{\prime} ; K_3 , K_2 , K_1 )
 \Bigr]_{ K^{\prime} = K_1^{\prime} + K_2^{\prime} - K}
 \Bigr\}
 \nonumber
 \\
 & & \hspace{-53mm } 
+ 9 \int_K   {\cal{S}}_{(1^{\prime},2^{\prime}),3^{\prime}}
 {\cal{S}}_{3, (2,1) }
 \Bigl[
 \bigl[ \dot{G}_{\Lambda} ( K )
 G_{\Lambda } ( K^{\prime} )
 +
 G_{\Lambda} ( K )
\dot{G}_{\Lambda} ( K^{\prime} )
 \bigr]
 \nonumber
 \Gamma^{(4)}_{ \Lambda }
 ( K_3^{\prime} , K^{\prime} ; K , K_3 )
 \Gamma^{(6)}_{\Lambda }
 ( K_1^{\prime} , K_2^{\prime}, K ; K^{\prime} , K_2 , K_1 )
  \Bigr]_{ K^{\prime} = K_3 - K_3^{\prime} +K }
 \nonumber
 \\
  & & \hspace{-53mm } + 9   \int_K
  {\cal{S}}_{(1^{\prime},2^{\prime}),3^{\prime}}
 {\cal{S}}_{3, ( 2, 1 )}
 \Bigl[  \bigl[
\dot{G}_{\Lambda} ( K )
 G_{\Lambda} ( K^{\prime} )
 G_{\Lambda} ( K^{\prime \prime} )
+
{G}_{\Lambda} ( K )
 \dot{G}_{\Lambda } ( K^{\prime} )
 G_{\Lambda } ( K^{\prime \prime} )
 + {G}_{\Lambda} ( K )
G_{\Lambda } ( K^{\prime} )
 \dot{G}_{\Lambda } ( K^{\prime \prime} )
\bigr]
 \nonumber
 \\
 & & \hspace{-31mm}
 \times
 \Gamma^{(4)}_{ \Lambda  }
 ( K_1^{\prime} , K_2^{\prime} ; K , K^{\prime} )
 \Gamma^{(4)}_{\Lambda }
 ( K_3^{\prime} , K^{\prime} ; K^{\prime \prime} , K_3 )
 \Gamma^{(4)}_{ \Lambda }
 ( K^{\prime \prime} , K ; K_2 , K_1 )
 \Bigr]_{ K^{\prime} = K_1^{\prime} + K_2^{\prime}
  - K}^{ K^{\prime \prime} = K_1 +
  K_2 - K}
 \nonumber
 \\
  & & \hspace{-53mm } + 36      \int_K
{\cal{S}}_{1^{\prime},  2^{\prime}, 3^{\prime} } {\cal{S}}_{1,2,3}
 \Bigl[
 \dot{G}_{\Lambda} ( K )
 G_{\Lambda } ( K^{\prime} )
 G_{\Lambda } ( K^{\prime \prime} )
 \nonumber
 \Gamma^{(4)}_{ \Lambda  }
 ( K_1^{\prime} , K^{\prime} ; K , K_1 )
 \Gamma^{(4)}_{\Lambda  }
 ( K_2^{\prime} , K^{\prime \prime} ; K^{\prime} , K_2 )
 \Gamma^{(4)}_{\Lambda }
 ( K_3^{\prime} , K ; K^{\prime \prime} , K_3 )
 \Bigr]_{ K^{\prime} = K_1 -  K_1^{\prime}
 + K}^{ K^{\prime \prime} = K_3^{\prime}
 - K_3 + K}
 \; .
 \nonumber
 \\
 & &
 \label{eq:sixpoint}
 \end{eqnarray}
 }
\end{widetext}

\begin{figure}
\epsfxsize8.9cm
\epsfbox{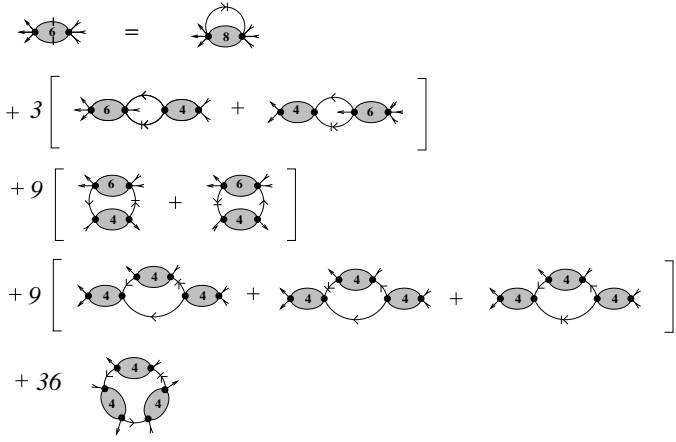}
\vspace{5mm}
\caption{
Diagrammatic representation of the flow equation
for the six-point vertex, see Eq.~(\ref{eq:sixpoint}).
}
\label{fig:SixPoint}
\end{figure}

\begin{figure}
\epsfxsize8.5cm
\hspace{5mm}
\epsfbox{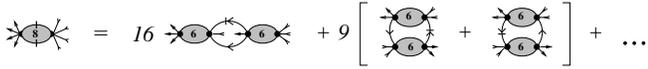}
\vspace{5mm}
\caption{
Diagrammatic representation of the terms of the flow equation
for the eight-point vertex which are of order $v_l^2$, 
see Eq.~(\ref{eq:eightpoint}).  The omitted
terms are at least cubic in the parameters $a_l$,$b_l$,$v_l$ and
$u_l$.
}
\label{fig:EightPoint}
\end{figure}

\subsection{Eight-point vertex}

In Eq.~(\ref{Gamma6punkt}) we also need
the following 
terms from the inhomogeneity of the eight-point vertex,
\begin{widetext}
 {\small
 \begin{eqnarray}
 &&
 \partial_\Lambda \Gamma^{(8)}_{\Lambda }
 ( K_1^{\prime} , K_2^{\prime} , K_3^{\prime} , K_4^{\prime} ;
 K_4 , K_3 , K_2  , K_1 ) 
\nonumber
\\
&&
\hspace{.2cm}
  =   16  \int_K  {\cal{S}}_{(1,2,3),4}{\cal{S}}_{
(1^\prime,2^\prime,3^\prime),4^\prime}
 \Bigl[
 \dot{G}_{\Lambda} ( K )   G_{\Lambda} ( K^{\prime} )
  \Gamma^{(6)}_{ \Lambda }
 ( K_1^{\prime} , K_2^{\prime}, K_3^{\prime} ;
 K_4 , K^{\prime} , K )  
  \Gamma^{(6)}_{ \Lambda }
 ( K, K^\prime,K_4^{\prime};
 K_3 , K_2 , K_1 )
 \Bigr]_{ K^{\prime} = K_1^\prime+K_2^\prime+K_3^\prime -K_4-K}
\nonumber 
\\
&& \hspace{.6cm}
 + 9 \int_K  {\cal{S}}_{ (1^{\prime},2^{\prime}),(3^{\prime},4^\prime)}
 {\cal{S}}_{ (1,2),(3,4)}
 \Bigl[
 \bigl[ \dot{G}_{\Lambda} ( K )
 G_{\Lambda } ( K^{\prime} )
 +
 G_{\Lambda} ( K )
\dot{G}_{\Lambda} ( K^{\prime} )
 \bigr]
 \nonumber 
 \\
 & &
\hspace{15mm}
 \times \
 \Gamma^{(6)}_{ \Lambda }
 ( K_3^\prime,K_4^{\prime} , K^{\prime} ; K , K_4,K_3 )
 \Gamma^{(6)}_{\Lambda }
 ( K_1^{\prime} , K_2^{\prime}, K ; K^{\prime} , K_2 , K_1 )
  \Bigr]_{ K^{\prime} = K_1^\prime+K_2^\prime-K_1-K_2+K } 
\;
+ \; \dots \; ,
 \label{eq:eightpoint}
 \end{eqnarray}
 }
\end{widetext}
with
\begin{subequations}
\begin{eqnarray}
{\cal S}_{(1,2,3),4}f(1,2,3,4)&=&\frac{1}{4}[f(1,2,3,4)+f(4,2,3,1)
 \nonumber \\ && \hspace{-9mm}
+f(1,4,3,2)
+f(1,2,4,3)] \; , 
\label{A1234a}
\end{eqnarray}
\begin{eqnarray}
{\cal S}_{(1,2),(3,4)}f(1,2,3,4)&=&\frac{1}{3}[f(1,2,3,4)+f(1,3,2,4)
 \nonumber \\ && \hspace{3mm} +f(1,4,2,3)] \; . \label{A1234b}
\end{eqnarray}
\end{subequations}

\section{Some integrals entering the contribution from
irrelevant terms to $\eta$}
\label{App:eta}
We here calculate the irrelevant contributions to the flow
of $\eta$, using Eq.~(\ref{eq:etaPSA}) of Sec.~\ref{subsect:etairr}.
Let us define
 \begin{eqnarray}
  \dot{f}_l (
  {\bf{q}}_1 ,   {\bf{q}}_2 )
 & = &
\frac{1}{1+r_l}
\Big<
 \frac{
\Theta ( 1 <  |  {\bf{q}}_1 + {\bf{q}}_2 - {\bf{\hat{q}}} | < e^{ l}
) }{
   |  {\bf{q}}_1 + {\bf{q}}_2 - {\bf{\hat{q}}} |^2 + r_l   }
 \nonumber
 \\
 & &
 \times
 \bigl[
 | {\bf{\hat{q}}} - {\bf{q}}_1 |
 +  | {\bf{\hat{q}}} - {\bf{q}}_2 | \bigr]
\Big>_{\bf \hat{q}}
 \label{eq:dotfldef}
 \; ,
 \end{eqnarray}
 \begin{eqnarray}
  \dot{g}_l (
  {\bf{q}}_1 ,   {\bf{q}}_2 )
 & = &
\frac{1}{1+r_l}
\Big<
 \frac{
\Theta ( 1 <  |  {\bf{q}}_1 + {\bf{q}}_2 - {\bf{\hat{q}}} | < e^{ l}
) }{|  {\bf{q}}_1 + {\bf{q}}_2 - {\bf{\hat{q}}} |^2 + r_l }
 \nonumber
 \\
 & &
 \times
 \bigl[
 | {\bf{\hat{q}}} - {\bf{q}}_1 |^2
 +  | {\bf{\hat{q}}} - {\bf{q}}_2 |^2 \bigr] \Big>_{\bf \hat{q}}
 \label{eq:dotgldef}
 \; ,
 \end{eqnarray}
\begin{eqnarray}
  \dot{h}_l (
  {\bf{q}}_1 ,   {\bf{q}}_2 )
 & = &
\frac{2}{1+r_l}
\Big<
 \frac{
\Theta ( 1 <  |  {\bf{q}}_1 + {\bf{q}}_2 - {\bf{\hat{q}}}|  < e^{ l} ) }
{|  {\bf{q}}_1 + {\bf{q}}_2 - {\bf{\hat{q}}} |^2 + r_l   }
 \nonumber
 \\
 & &
 \times
 | {\bf{\hat{q}}} - {\bf{q}}_1 |
  | {\bf{\hat{q}}} - {\bf{q}}_2 |
\Big>_{\bf \hat{q}}
 \label{eq:dothldef}
 \; .
 \end{eqnarray}

To calculate the contributions of the six-point vertex to
Eq.~(\ref{eq:etaPSA}), we need to evaluate (we restrict the results
to $D=3$)

\begin{widetext}
\begin{eqnarray}
\frac{\partial}{\partial { q}^2} \left< \dot{\Gamma}_l^{6}(\lambda
{\bf q},\lambda {\bf \hat{q}^\prime},\lambda^\prime {\bf
\hat{q}^{\prime\prime}};\lambda^\prime {\bf \hat{q}^{\prime\prime}}, \lambda {\bf
\hat{q}^\prime}, \lambda {\bf q}) \right>_{{\bf \hat{q}^\prime},{\bf \hat{q}^{\prime\prime}}}
\Big|_{{\bf q}^2=0} =
\nonumber \\ && \hspace{-7cm}
 -v_l \Big\{ 3 u_l \left[ \dot{\chi}^{\prime
\prime}_l(\lambda,\lambda)+ \dot{\chi}^{\prime
\prime}_l(\lambda,\lambda^\prime) \right] 
+2(b_l+2 a_l) \big[\dot{f}^{\prime
\prime}_l(\lambda,\lambda)+ \dot{f}^{\prime
\prime}_l(\lambda,\lambda^\prime) +
\dot{\alpha}_l(\lambda,\lambda^\prime) \big]+ (b_l +a_l)
\dot{\varphi}^{\prime \prime}_l(\lambda) \Big\} \; ,
\end{eqnarray}
with
\begin{eqnarray}
  \dot{f}_l^{\prime \prime}(\lambda,\lambda^\prime)&=&
  \frac{\partial}{\partial  q^2}
  \left. \left< \dot{f}_l\left( \lambda \bf{q},
        \lambda^\prime \bf{\hat{q}^\prime}\right) \right>_{\bf \hat{q}^\prime} \right|_{q^2=0}
  =\frac{-\lambda^2(1+\lambda^\prime)^2+\lambda^2(3+2\lambda^\prime) r_l}
  {12\lambda^\prime (1+r_l)\left[
      (1+\lambda^\prime)^2+r_l\right]^2} \; ,
  \\
  \dot{\varphi}_l^{\prime \prime}(\lambda) & = &
  \dot{\chi}_l(0)
  \frac{\partial}{\partial q^2} \left. \Big<
    |{\bf \hat{q}^\prime}+\lambda \bf{q}|
    \Big>_{\bf \hat{q}^\prime}
  \right|_{q^2=0} = \frac{\lambda^2 }{3 (1+r_l)^2} \; ,
  \\
  \dot{\chi}_l^{\prime \prime}(\lambda,\lambda^\prime) & = &
  \frac{2}{1+r_l}  \frac{\partial}{\partial  q^2}
  \left. \left< \tilde{G}_l\left( \lambda \bf{q}+
        \lambda^\prime \bf{\hat{q}^{\prime\prime}} +\bf{\hat{q}^\prime}
      \right) \right>_{{\bf \hat{q}^\prime},{\bf \hat{q}^{\prime\prime}}} \right|_{q^2=0}
  =\frac{\lambda^2 [r_l-(1+\lambda^\prime)^2]}{6 (1+r_l) \lambda^\prime
    [(1+\lambda^\prime)^2+r_l]^2} \; ,
  \\
  \dot{\alpha}_l(\lambda,\lambda^\prime) & = &
  \frac{1}{1+r_l}  \frac{\partial}{\partial  q^2}
  \left. \left< \tilde{G}_l\left( \lambda \bf{q}+
        \lambda^\prime \bf{\hat{q}^{\prime\prime}} +\bf{\hat{q}^\prime}
      \right)
      | \lambda \bf{q}+\lambda^\prime \bf{\hat{q}^{\prime\prime}}|
    \right>_{{\bf \hat{q}^\prime},{\bf \hat{q}^{\prime \prime}}} \right|_{q^2=0}
  =\frac{\lambda^2
    [(1+\lambda^\prime)^2(2+\lambda^\prime)+(2+3\lambda^\prime)r_l]}
  {12(1+r_l)\lambda^\prime [(1+\lambda^\prime)^2+r_l]}\; .
\end{eqnarray}
To calculate the contributions of the BCS, ZS and ZS$^\prime$
channel to Eq.~(\ref{eq:etaPSA}), further averages are required:

\begin{eqnarray}
&& 
\hspace{-.9cm}
\frac{\partial}{\partial q^2} \left. \left< \dot{g}_l\left(
\lambda \bf{q}, \lambda \bf{\hat{q}^\prime}\right) \right>_{\bf \hat{q}^\prime}
\right|_{q^2=0} =\frac{-\lambda \left[
\lambda(1+\lambda)^2(2+3\lambda)+(-4+(-2+\lambda)
\lambda) r_l+2 ((1+\lambda)^2+r_l)^2 
\ln \left(\frac{1+r_l}{(1+\lambda)^2+r_l}\right)
\right]}{12 (1+r_l)\left[(1+\lambda)^2+r_l\right]^2} \; , 
\end{eqnarray}
\begin{eqnarray}
&& \hspace{-.9cm}
\frac{\partial}{\partial q^2} \left. \left< \dot{f}_l\left(
\lambda \bf{q}, \lambda \bf{\hat{q}^\prime}\right) | \lambda \bf{q}+ \lambda
\bf{\hat{q}^\prime}|\right>_{\bf \hat{q}^\prime} \right|_{q^2=0} =
\frac{\lambda\left[8(1+r_l)+\lambda\left[(3+\lambda)^2(
3+\lambda(2+\lambda))+2
(10+\lambda(6+\lambda))r_l+r_l^2\right]\right] } {24
(1+r_l)\left[(1+\lambda)^2+r_l\right]^2} 
\nonumber 
\\ && 
\hspace{-.9cm}
 \hspace{5mm}+ \frac{
2(1-\lambda^2+r_l)\left[(1+\lambda)^2+r_l\right]^2
\mbox{arctanh}\left( \frac{\lambda
\sqrt{-r_l}}{1+\lambda+r_l}\right)+(1+\lambda^2
+r_l)\left[(1+\lambda)^2+r_l\right]^2 \sqrt{-r_l}
\ln \left(\frac{1+r_l}{(1+\lambda)^2+r_l}\right) } {24
(1+r_l)\left[(1+\lambda)^2+r_l\right]^2 \sqrt{-r_l}} \; ,
\\
%
&& \hspace{-.9cm}
\frac{\partial}{\partial q^2} \left. \left< \dot{h}_l\left(
\lambda \bf{q}, \lambda \bf{\hat{q}^\prime}\right) \right>_{\bf \hat{q}^\prime}
\right|_{q^2=0}=
 \nonumber 
\\ && \hspace{-.5cm}
\hspace{7mm}
-\lambda\frac{
\left[(\lambda-1)(1+\lambda)^2
\left(\lambda+\lambda^2+2r_l\right) +\lambda r_l^2 \right] \sqrt{-r_l}
+(1-\lambda^2-r_l)\left[(1+\lambda)^2+r_l\right]^2\mbox{arctanh}\left(
\frac{\lambda \sqrt{-r_l}}{1+\lambda+r_l}\right) } {6
(1+r_l) \left[(1+\lambda)^2+r_l\right]^2 \sqrt{-r_l}} \; ,
\end{eqnarray}
\begin{eqnarray}
&& \hspace{-.5cm}
\dot{\chi}_l(0) \frac{\partial}{\partial q^2} \left.  \big<
|{\bf \hat{q}^{\prime\prime}}+\lambda \bf{q}| \ |{\bf \hat{q}^{\prime\prime}}+\lambda
\bf{\hat{q}^\prime}| \big>_{{\bf \hat{q}^{\prime}}, \bf \hat{q}^{\prime\prime}}
\right|_{q^2=0} = \frac{\lambda^2 (3+\lambda^2)}{9 (1+r_l)^2} \; ,
\\ 
&& \hspace{-.5cm}
\frac{\partial}{\partial q^2} \left. \Big< \dot{\chi}_l(|\lambda{\bf
\hat{q}^\prime}+\lambda \bf{q}|) \ |\lambda{\bf \hat{q}^\prime}+\lambda \bf{q}|^2
\Big>_{\bf \hat{q}^\prime} \right|_{q^2=0}= 
\nonumber \\ &&
\hspace{-.5cm}
\hspace{5.2cm}
\frac{\lambda^2(1+\lambda)^2(4+3\lambda) +\lambda^2(4+5\lambda)
r_l + \lambda\left[(1+\lambda)^2+r_l \right]^2
\ln \left(
  \frac{(1+\lambda)^2+r_l}{1+r_l}
  \right)
}{6 (1+r_l)\left[(1+\lambda)^2+r_l \right]^2} 
\; .
\end{eqnarray}
Using these expressions to calculate $\dot{\Gamma}_\infty^{(4)}$ from
Eqs.~(\ref{Gamma6punkt},\ref{sixpointscale},\ref{DotGamma6zero},\ref{eq:BCSfull},\ref{eq:ZSfull}) 
one can 
determine $\eta$ via  Eq.~(\ref{eq:etaPSA}).
One can check
 that to linear order in $\lambda$ one has
\begin{equation}
\frac{\partial}{\partial {q^2}}\big<\dot{\Gamma}^{(4)}_l(\lambda{\bf q},\lambda{\bf
\hat{q}^\prime},\lambda {\bf \hat{q}^\prime},\lambda{\bf q})\big>_{\bf \hat{q}^\prime}
|_{q^2=0}\simeq\lambda (\partial_la_l+\partial_lb_l)/3+ 2 \lambda
(a_l+b_l) \eta_l/3 \; ,
\end{equation}
 which, after performing the
$\lambda$-integration, reproduces at the fixed point the result
Eq.~(\ref{eq:etares}).

\end{widetext}

\end{appendix}

\end{document}